\documentclass[twocolumn,twocolappendix]{aastex7}
\usepackage{natbib}
\usepackage{comment}
\usepackage{amsmath}
\usepackage{algorithm}
\usepackage{amsfonts}
\usepackage{mathrsfs}
\usepackage{amssymb}
\usepackage{color}
\usepackage{graphicx} 
\let\tablenum\relax
\usepackage{siunitx}

\newcommand{\nus}{\textit{NuSTAR} }
\newcommand{\xmm}{\textit{XMM-Newton} }

\newcommand{\myt}{\texttt{MYTorus }}
\newcommand{\uxc}{\texttt{UXClumpy} }
\newcommand{\rxt}{\texttt{RXTorusD} }

\newcommand{\ctk}{\texttt{CTKcover}}
\newcommand{\tor}{\texttt{TORsigma}}
\newcommand{\sil}{\citetalias{Silver2023} }

\begin{document} 

\defcitealias{Silver2023}{S23}
\defcitealias{Asmus2015}{A15}

\correspondingauthor{Ross Silver}
\author[0000-0001-6564-0517]{Ross Silver}
\email{rosssilver.astro@gmail.com}
\affiliation{NASA Goddard Space Flight Center, Greenbelt, MD 20771, USA}
\affiliation{Southeastern Universities Research Association, Washington, DC 20005, USA}
\affiliation{School of Physics and Astronomy, University of Minnesota, Minneapolis, MN 55455, USA}

\author[0000-0003-3638-8943]{N\'{u}ria Torres-Alb\`{a}} \email{nuria@virginia.edu}
\affiliation{Department of Astronomy, University of Virginia, P.O. Box 400325, Charlottesville, VA 22904, USA}
\affiliation{Department of Physics and Astronomy, Clemson University,  Kinard Lab of Physics, Clemson, SC 29634, USA}

\author[0000-0001-5544-0749]{Stefano Marchesi} \email{stefano.marchesi@inaf.it}
\affiliation{Dipartimento di Fisica e Astronomia (DIFA), Università di Bologna, via Gobetti 93/2, I-40129 Bologna, Italy}
\affiliation{Department of Physics and Astronomy, Clemson University,  Kinard Lab of Physics, Clemson, SC 29634, USA}
\affiliation{INAF - Osservatorio di Astrofisica e Scienza dello Spazio di Bologna, Via Piero Gobetti, 93/3, 40129, Bologna, Italy}

\author[0000-0002-9719-8740]{Vittoria E. Gianolli} \email{vgianol@clemson.edu}
\affiliation{Department of Physics and Astronomy, Clemson University,  Kinard Lab of Physics, Clemson, SC 29634, USA}

\author[0000-0003-2287-0325]{Isaiah Cox} \email{isc@clemson.edu}
\affiliation{Department of Physics and Astronomy, Clemson University,  Kinard Lab of Physics, Clemson, SC 29634, USA}

\author[0009-0002-6991-1534]{Dhrubojyoti Sengupta} \email{dsengup@clemson.edu}
\affiliation{Center for Space Science and Technology, University of Maryland Baltimore County, 1000 Hilltop Circle, Baltimore, MD 21250, USA}

\author[0000-0002-7825-1526]{Indrani Pal} \email{ipal@clemson.edu}
\affiliation{Department of Physics and Astronomy, Clemson University,  Kinard Lab of Physics, Clemson, SC 29634, USA}

\author[0000-0002-6584-1703]{Marco Ajello} \email{majello@clemson.edu}
\affiliation{Department of Physics and Astronomy, Clemson University,  Kinard Lab of Physics, Clemson, SC 29634, USA}

\author[0000-0002-7791-3671]{Xiurui Zhao} \email{xiurui.zhao.work@gmail.com}
\affiliation{Cahill Center for Astrophysics, California Institute of Technology, 1216 East California Boulevard, Pasadena, CA 91125, USA}

\author[0009-0003-3381-211X]{Kouser Imam} \email{kimam@clemson.edu}
\affiliation{Department of Physics and Astronomy, Clemson University,  Kinard Lab of Physics, Clemson, SC 29634, USA}

\author[0000-0001-7796-8907]{Anuvab Banerjee} \email{anuvabb@clemson.edu}
\affiliation{Department of Physics and Astronomy, Clemson University,  Kinard Lab of Physics, Clemson, SC 29634, USA}

%\author[0000-0002-2115-1137]{Francesca Civano} %\email{francesca.m.civano@nasa.gov}
%\affiliation{NASA Goddard Space Flight Center, Greenbelt, MD 20771, USA}

   \title{Compton-thick AGN in the NuSTAR Era. XI. Analyzing 11 CT-AGN Candidates Selected with Machine Learning }

%
%-------------------------------------------------------------------

\begin{abstract}
    This work discusses the broadband X-ray spectral analysis of 11 candidate heavily-obscured active galactic nuclei (AGN) selected based on their infrared and X-ray properties by a recently published machine learning algorithm. This paper is part of a larger work to identify and characterize all AGN in the local universe ($z < 0.1$) with the largest line-of-sight (los) column densities (N$_{\rm H}$), the so-called Compton-thick (CT-, N$_{\rm H, los} \geq$ 10$^{24}$ cm$^{-2}$) AGN. We modeled the X-ray spectra using two physically-motivated models, \uxc and \texttt{RXTorusD}. Of the 11 AGN in our sample, we found three to be obscured with 22.7 $<$ LogN$_{\rm H, los} \leq$ 23.0, five have 23.0 $<$ LogN$_{\rm H, los} \leq$ 23.25, and three have 23.4 $<$ LogN$_{\rm H, los} \leq$ 23.9, according to \texttt{UXClumpy}. Meanwhile, according to \texttt{RXTorusD}, we found three AGN to be obscured with 22.7 $<$ LogN$_{\rm H, los} \leq$ 23.0, four with 23.0 $<$ LogN$_{\rm H, los} \leq$ 23.4, and four with 23.85 $<$ LogN$_{\rm H, los} \leq$ 23.96. %Based on the upper uncertainty values from \texttt{RXTorusD}, three sources are potentially CT. 
    Additionally, this work served as a comparison between \uxc and \texttt{RXTorusD}. We found broad agreement between the two, with 8/11 sources agreeing on the value of the photon index $\Gamma$, while only 5/11 sources agreeing on the N$_{\rm H, los}$ value within the 90\% confidence level.
\end{abstract}

\section{Introduction} \label{sec:intro}
Supermassive black holes (SMBHs; log$M_{BH}$/$M_{\odot}$ $\sim$ 6 -- 9.5) that accrete surrounding material and emit throughout the entire electromagnetic spectrum are known as active galactic nuclei (AGN). AGN are one of the most powerful, consistent emitters in the Universe. Numerous works have shown a strong correlation between the SMBH and the properties of the host galaxy they reside in, such as the host galaxy bulge, velocity dispersion, and luminosity \citep[e.g.,][]{Magorrian_1998, Richstone1998, Gebhardt_2000, Merritt_2001, Ferrarese2005, Kormendy2013}. From these trends, it can be inferred that SMBHs play a significant role in determining stellar formation rates through feedback \citep[e.g.,][]{Ferrarese2000, Gebhardt_2000, DiMatteo2005, Merloni_2010, Fiore2017, MartinNavarro2018}. This suggests that the growth of AGN has a strong link with the development of the host galaxy. Therefore, understanding how AGN accrete and change over time is crucial for understanding the co-development of SMBHs and their host galaxies. \\
\indent One of the best tools we have to study the evolution of AGN over cosmic time is the cosmic X-ray background \citep[CXB;][]{Giacconi1962}, i.e., the diffuse emission covering the entire sky in the $E \sim$ 1--300\,keV band \citep{Alexander_2003, Gilli2007, Treister_2009, Ueda_2014, Brandt2021CXB}. The CXB is produced primarily by AGN and therefore, holds the secrets to their growth and evolution within its spectrum. This information can be uncovered by developing population synthesis models, like those of \cite{Ueda_2014} and \cite{Ananna2019}. These models
attempt to reproduce the CXB by assuming various properties, such as the distribution of AGN with different levels of obscuration, known as the line-of-sight column density (N$_{\rm H, los}$). In particular, these population synthesis models uncovered a large discrepancy regarding the most obscured subclass of AGN, the so-called Compton-thick (CT-) AGN, with N$_{\rm H, los} \geq$ 10$^{24}$ cm$^{-2}$. While current works have found an observed CT-fraction ranging from 5-20\% in the local universe \citep[$z < 0.01$;][]{Burlon2011, Ricci_2015, TorresAlba2021}, population synthesis models \citep{Ueda_2014, Ananna2019} predict the CT-fraction to be as high as 50\% of all AGN. This is likely due to the fact that the shape and magnitude of the peak of the CXB \cite[$\sim$30\,keV;][]{Ajello_2008} can only be reproduced by a substantial, yet-to-be-fully-detected, population of CT-AGN \citep{Gilli2007, Ananna2019}. \\
\indent Due to their extreme levels of obscuration, CT-AGN block much of the emission from the central regions (particularly in the optical through the soft X-rays), thus making them more difficult to detect. Therefore, creative methods must be developed to close the gap between the observed and predicted fraction of CT-AGN. To this end, \citep[][henceforth \citetalias{Silver2023}]{Silver2023}, developed a multiple linear regression machine learning (ML) algorithm designed to accurately predict the line-of-sight column densities of AGN using data from the infrared, soft and hard X-rays. The input parameters used to train the algorithm included 6 WISE colors, a ratio between the 12$\mu$m and 2$-$10\,keV fluxes, two soft X-ray hardness ratios (HRs), and hard X-ray count rates from \textit{Swift}-BAT \citep{Gehrels2004}. These parameters have all been shown to be reliable predictors of column densities (see references within \citetalias{Silver2023}). The training and testing sample consisted of 451 AGN with column densities determined by spectral fitting of soft X-ray spectra with hard X-ray spectra from \textit{Swift}-BAT, ranging from LogN$_{\rm H, los}$ = 20.14 -- 25.0. This method found a significant reduction in false positives (sources predicted to have N$_{\rm H, los} \geq$ 10$^{23}$, but had a true value of N$_{\rm H, los} \leq$ 10$^{22}$) compared with a single mid-infrared (MIR) $-$ X-ray flux (or luminosity) ratio from other published works \citep{Asmus2015, Pfeifle2022}. \\
\indent The purpose of the method in \sil is to find the most promising heavily obscured AGN candidates and then observe them with the X-ray instruments XMM-\textit{Newton} \citep{Jansen2001} and the Nuclear Spectroscopic Telescope Array \citep[\textit{NuSTAR};][]{Harrison2013}. The combination of the superb effective area of \xmm around the iron line (5--7\,keV) and the sensitivity of \nus above 10\,keV make this the ideal method to accurately determine the X-ray column density of an AGN, as well as accurately measure other important properties of the obscuring medium, such as its covering factor and average column density, among others. In this work, we present for the first time the X-ray spectral fitting results of 11 AGN found to be promising heavily obscured AGN candidates from the algorithm presented in \citetalias{Silver2023}. This paper is organized as follows: Section \ref{sec:data_reduc} lays out the data reduction process for both \xmm and \textit{NuSTAR}. Section \ref{sec:analysis, models} describes the two physically-motivated models used in this work. Section \ref{sec:fitting_results} discusses the general fitting results for the entire sample and uses them to make comparisons. Section \ref{sec:summ_+_conc} summarizes the paper and next steps in our work. Appendix \ref{app_sec:ind_source} discusses the fitting results for each individual source and Appendix \ref{app_sec:tab+sepc} displays the best-fit parameters and spectra. \\
\indent The uncertainties listed in this paper are at the 90\% confidence level unless mentioned otherwise. We adopt standard flat cosmological parameters: H$_0$ = 70 km s$^{-1}$ Mpc$^{-1}$, q$_0$ = 0.0, and $\Omega_\Lambda$= 0.73.

\section{Observations and Data Reduction} \label{sec:data_reduc}
The 11 galaxies analyzed in this work were observed as part of two \nus proposals (proposal ID 9093, PI: Silver, and proposal ID 10306, PI: Silver) dedicated to identifying CT-AGN using the machine learning algorithm first described in \citetalias{Silver2023}. As stated above, \textit{Swift}-BAT is an all-sky X-ray instrument and is the least biased instrument towards obscured sources. Therefore, 
our sample selection started with sources detected in the \textit{Swift}-BAT 150-month catalog\footnote{The catalog can be accessed here: \url{https://science.clemson.edu/ctagn/bat-150-month-catalog/}} (Imam et al. submitted), which when the \citetalias{Silver2023} algorithm is applied to them, yield a line-of-sight column density logN$_{\rm H} >$ 23.8. For the proposals, we selected 11 sources that did not have previous \nus observations. Of the 11 sources, eight were granted quasi-simultaneous ($\leq$ 1 day apart) observations with \xmm (three from cycle 9 and all five from cycle 10). A summary of all observations is shown in Table \ref{tab:observations}. \\
\indent The \nus data were derived from the two focal plane modules, FPMA and FPMB. The raw data files were screened, calibrated, and cleaned using \texttt{nupipeline} version 0.4.9. We used the \nus calibration database version 20250218 alongside \texttt{nuproducts} to create the redistribution matrix function, the auxiliary response function, and the light-curve files. We extracted the source spectra using a 50$\arcsec$ circular region and the background spectra using an annulus with inner radius of 100$\arcsec$ and outer radius of 150$\arcsec$. When placing the background region, the images were visually inspected to ensure the spectra would not be contaminated by any nearby sources. Then, the HEAsoft\footnote{\url{https://heasarc.gsfc.nasa.gov/docs/software/heasoft/}} tool \texttt{grppha} was used to group each spectra with at least 15 counts per bin. \\ %However, in the case of UGC 12566, there were too few counts and thus the spectra was binned to 15 counts per bin. 
\indent The \xmm observations were reduced using version 21.0.0 of the Science Analysis System \citep[SAS;][]{Jansen2001}. We removed all flaring periods for each observation, which resulted in an average of $\sim$30\% loss of exposure time per observation. Once this was completed, we extracted the source spectra using a 15$\arcsec$ circular region. The background spectra were extracted using an annulus with inner radius of 75$\arcsec$ and outer radius of 100$\arcsec$. Once again, the images were visually inspected to prevent any contamination from nearby sources in the background spectra. The data from all three modules (MOS1, MOS2, and PN) were jointly fit with linked parameters, assuming minimal cross-calibration uncertainties between the three modules. Each spectra was then binned with at least 15 counts per bin.

\begingroup
\renewcommand*{\arraystretch}{1.4}
\begin{table*}
    %\centering
    \caption{Summary of \xmm and \nus Observations.}
    \label{tab:observations}
    \begin{tabular}{ccccccc}
    \hline \hline
    Source Name & Instrument & Sequence & Observation Date & $z$\footnote{Obtained from \url{https://ned.ipac.caltech.edu}} & Exp. & Net Ct. Rate \\
     &  & ObsID & (UTC) & & (ks) & 10$^{-2}$ cts s$^{-1}$ \\
     \hline 
     IGR J19118$-$1707 & \nus & 60960001002 & 2023-10-09 & 0.0236 & 45.9 & 4.4 \\
     & XMM & 930790101 & 2023-10-09 &  & 39.2 & 4.9 \\
    NGC 4250 & \nus & 60960002002 & 2023-10-05 & 0.0068 & 41.8 & 2.0 \\
    & XMM & 930790201 & 2023-10-05 & & 29.2 & 5.4 \\ 
    MCG--02--34--058 & \nus & 60960003002 & 2024-01-06 & 0.0216  & 34.5 & 5.0 \\
    & XMM & 930790301 & 2024-01-07 & & 32.0 & 6.8  \\
    2MASX J18454978$-$5548252 & \nus & 60960004002 & 2023-08-10 & 0.0343 & 32.6 & 4.5 \\
    \\
    2MASX J21570549$+$0632169 & \nus & 60960005002 & 2023-06-21 & 0.0306 & 30.2 & 12.9 \\
    \\
    SDSS J073323.77$+$441448.8 & \nus & 60960006002 & 2023-10-11 & 0.0559 & 28.8 & 4.3 \\
    \\
    \hline 
    2MASX J13100789$-$1711398 & \nus & 61065002002 & 2025-01-03 & 0.0262 & 27.3 & 4.4 \\
    & XMM & 0952020201 & 2025-01-04 & & 29.0 & 2.5 \\
    LEDA 2019751 & \nus & 61065003002 & 2024-11-04 & 0.0826 & 38.9 & 6.2 \\
    & XMM & 0952020301 & 2024-11-04 & & 39.8 & 8.7 \\
    NGC 52 & \nus & 61065001002 & 2025-01-03 & 0.0180 & 37.9 & 7.6 \\
    & XMM & 0952020101 & 2025-01-03 & & 43.8 & 6.7 \\
    UGC 12566 & \nus & 61065005002 & 2025-01-09 & 0.0193 & 20.7 & 0.7 \\
    & XMM & 0952020501 & 2025-01-09 & & 31.4 & 0.5 \\
    NGC 6657 & \nus & 61065004002 & 2025-03-19 & 0.0233 & 26.8 & 6.2 \\
    & XMM & 0952020401 & 2025-03-19 & & 37.0 & 5.8 \\
    \hline \hline
    \end{tabular}
    \newline \textbf{Notes:} \\
Average count rate (in cts s$^{-1}$), weighted by the exposure for \xmm
and \textit{NuSTAR}, where observations from multiple instruments are combined. Count rates are computed in the 2$-$10\,keV and 3$-$70\,keV band, respectively. The horizontal dividing line separates the sources observed in \nus cycles 9 (top) and 10 (bottom).\\% The exposure times listed are before the removal of flares. \\

\end{table*}
\endgroup

\section{Spectral Models} \label{sec:analysis, models}
All of the spectral fitting was performed using \texttt{XSPEC} version 12.13.1. The Galactic column densities in the direction of each source were obtained with the Heasoft tool \texttt{nh} \citep{Kalberla05}. We calculated the intrinsic luminosity of each source using \texttt{clumin}\footnote{\url{https://heasarc.gsfc.nasa.gov/xanadu/xspec/manual/node285.html}}. In this section, we discuss the physically-motivated models used to analyze the X-ray spectra of our 11-source sample. \\
\indent We note that while the observations between \nus and \xmm were taken quasi-simultaneously, non-negligible flux variations between the telescopes can still occur. Therefore, we have included a cross-calibration constant (\textit{c$_{nus}$}) to account for this. We have also added the component \texttt{apec}\footnote{\url{http://astroa.physics.metu.edu.tr/MANUALS/xspec12_html/XSmodelApec.html}} to model the thermal bremsstrahlung emission that originates in the hot gas inside the host galaxy. \\
\indent For all sources with XMM observations, we modeled the data in the energy range 0.2--12\,keV. For  \textit{NuSTAR}, only data with an energy $>$3\,keV were kept. The maximum energy changed for each source depending on when the data become background-dominated. For example, MCG--02--34--058 is the source which becomes background-dominated at the lowest energies, E$\sim$30\,keV. The source which becomes background-dominated at the highest energies (E$\sim$70\,keV) is instead 2MASX J21570549$+$0632169.

\subsection{UXClumpy}
\uxc \citep{Buchner2019} is a widely used model \citep[e.g.,][]{Silver2022_XMM, Kayal2023, TorresAlba2023, Belvedersky2025, Cox2025}, which assumes a clumpy distribution of material. It is a physically motivated model that contains a distribution of clouds of varying sizes and densities, which can be placed in a broader or more compact distribution around the plane of the central engine. %The Monte Carlo X-ray radiative transfer code, XARS, is used within the model to calculate the X-ray spectra for obscured AGN. 
\uxc can be applied in \texttt{XSPEC} using the following formula:

\begin{eqnarray}
 \label{eq:uxc}
    ModelA = constant_1  *  phabs * \\  \nonumber (apec + uxcl\_\textit{cutoff}.fits + \\ f_s * uxcl\_\textit{cutoff}\_omni.fits). \nonumber
\end{eqnarray}

The first table, \textit{uxcl\_cutoff.fits}, represents the transmitted, reflected, and fluorescent line emission. The reflection component is dependent on the cloud distribution generated by \texttt{UXClumpy}. These are modeled with the inclination angle ($\theta_{inc}$ = 0--90$\degr$) and the cloud distribution dispersion \texttt{TORsigma} ($\sigma$ = 6--90$\degr$). In some instances when sources are severely reflection dominated, an additional Compton-thick reflector located near the corona can be added. This is modeled by \texttt{CTKcover}, the covering factor of the inner ring reflector, which ranges from \texttt{CTKcover} = 0--0.6. One can think of this as a thick material that blocks the primary emission from the corona, but also reflects it into our line of sight. \\
\indent The second table, \textit{uxcl\_cutoff\_omni.fits}, represents the elastically scattered emission that pierces through the torus.  This second power law was linked to the main one, such that the values of the photon index, normalization, and high-energy cutoff were equal to those from the primary X-ray continuum.  {The relative normalization of this second component with respect to the main one is parameterized by} the scattering fraction $f_s$, usually $<$10\% \citep{Ricci2017}, although it can be as high as 15--20\% \citep{Marchesi2016, Gupta2021}\footnote{We note that this only applies to \texttt{RXTorusD}, as \uxc states in its documentation that a scattering fraction above 10\% is not physical in this model \citep{Buchner2019}.}.\\

\subsection{RXTorusD}
The second model we used in this work is \rxt \citep{ricci2023}. Unlike \texttt{UXClumpy}, \rxt assumes a homogeneous geometry. It is similar to the geometry of the widely-used model \myt \citep{Murphy2009}, however \rxt allows the covering factor of the torus to vary, while this parameter is fixed to 0.5 in \texttt{MYTorus}. Moreover, \texttt{MYTorus} includes two constants that re-calibrate subcomponents of the reflection continuum and lines, and account for potential deviations between the real data and the model's geometry. The actual value of these constants, however, is not directly related to any physical quantity, which hinders interpretation. Even so, numerous works have shown that homogeneous models are capable of accurately modeling X-ray spectra \citep[see e.g.][]{Zhao2019a, Zhao2019b, TorresAlba2021, Silver2022, Kayal2023} and how the derived parameters, such as N$_{\rm H,los}$, tend to agree with those provided by the clumpy model \texttt{UXClumpy} \citep[e.g.][]{TorresAlba2023, Pizzetti2025, TorresAlba2025}. \\
\indent \rxt uses the Monte Carlo code \texttt{REFLEX}\footnote{https://www.astro.unige.ch/reflex/} \citep{Paltani2017} to self-consistently simulate the absorption and reprocessed emission in AGN spectra. The model can be implemented in \texttt{XSPEC} using the following:

\begin{eqnarray}
 \label{eq:rxt}
    ModelB = constant_1  *  phabs * \\   (apec + RXTorusD\_rprc.mod + \nonumber \\ RXTorusD\_cont.mod * \textit{zcutoffpl} + \nonumber \\ f_s * \textit{cutoffpl}). \nonumber
\end{eqnarray}

The first table accounts for all the reprocessed components, i.e., the reflection and fluorescence. The second table provides the absorption which is multiplied by a cutoff power law representing the primary continuum from the corona. The final component is a cutoff power law that is controlled by the scattering fraction $f_s$. For both \uxc and \texttt{RXTorusD}, the high-energy cutoff was set to 300\,keV. \\
\indent Unlike \texttt{UXClumpy}, \rxt includes two separate N$_{\rm H}$ values: a line-of-sight column density (N$_{\rm H,los}$) and an equatorial column density (N$_{\rm H,eq}$). The N$_{\rm H,los}$ is included in the direct continuum (\textit{cont.mod}) table while the N$_{\rm H,eq}$ is incorporated in the reprocessed component (\textit{rprc.mod}) table.
When decoupled, these parameters give \rxt the ability to imitate a clumpy model by providing two separate column density values. However, two sources (SDSS J073323.77$+$441448.8 and UGC 12566) lacked a strong enough reflection component to constrain this parameter. For these two cases, we fixed the N$_{\rm H,eq}$ to the value of the N$_{\rm H,los}$. We note that this equivalency is only valid in an edge-on geometry, since the N$_{\rm H,los}$ decreases with respect to the N$_{\rm H,eq}$ with decreasing inclination angle, moving away from the equivalency. Since the reflection component parameters could not be constrained for either of these two sources, we elect to assume this edge-on geometry by fixing the inclination angle to 90$\degr$ and setting N$_{\rm H,eq}$ equal to N$_{\rm H,los}$.\\
\indent \rxt also includes parameters to model the inclination angle ($\theta$ = 0--90$\degr$) and the ratio of the inner-to-outer radius of the torus (r/R = 0--1). For comparison, the geometry of \myt is fixed to r/R = 0.5. Both parameters are included in the reprocessed table shown in Equation \ref{eq:rxt}. \\
\indent Both models considered in this work have a maximum of ten free parameters in every fit. Eight of them are shared between models: the \texttt{apec} temperature and normalization, the main power law photon index and the normalization, the inclination angle of the torus, the line-of-sight column density, the scattering fraction, and the cross-calibration constant. The two parameters unique to \uxc are the \texttt{CTKcover} and \texttt{TORsigma}. The two parameters unique to \rxt are the ratio of the inner-to-outer radius of the torus and the equatorial column density. For both models, these sets of two parameters describe the geometry of the reflecting material.

\section{Fitting Results and Discussion} \label{sec:fitting_results}

The following sections will discuss details about our fitting process and then contextualize the fitting results, and more specifically, focus on two comparisons: 1) the predicted N$_{\rm H,los}$ values versus the best-fit model values, and 2) the best-fit N$_{\rm H,los}$ and photon indices $\Gamma$ from \uxc compared with those found by \texttt{RXTorusD}.

\begin{figure}
    \hspace{-0.8cm}
    \includegraphics[scale=0.6]{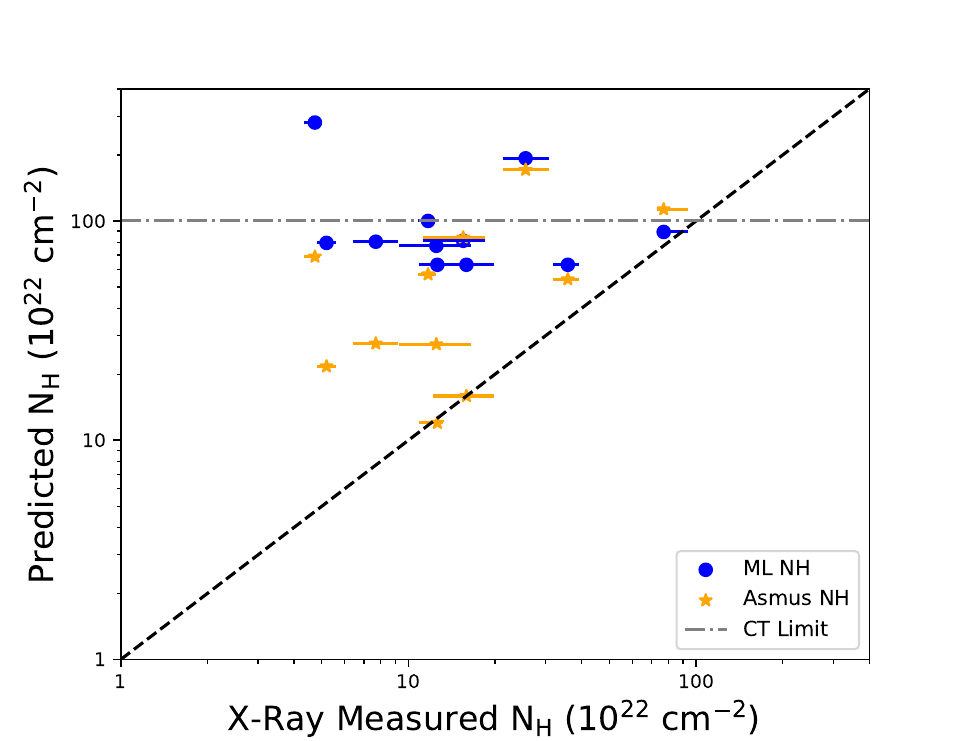}
    \caption{The line-of-sight column density for all 11 sources predicted by two different techniques versus the best-fit \uxc N$_{\rm H,los}$ presented in this work. The blue circles represent the values predicted by the machine learning algorithm presented in \cite{Silver2023} and the orange stars are the N$_{\rm H,los}$ values predicted by the relation in \cite{Asmus2015}. The dashed diagonal line represents the 1:1 relationship between the predicted and X-ray-measured values. The dash-dotted horizontal line represents the boundary where any source above it was predicted to be CT by the \citetalias{Silver2023} algorithm.}
    \label{fig:xr_vs_ml}
\end{figure}

\subsection{Fitting Details}

In our modeling, we left the \nus cross-calibration constant $c_{nus}$ free to vary. Typically, the \nus flux normalization is about 10-20\% higher than XMM \citep[see Table 6;][]{Madsen2017}, which is consistent with what was found in this work. \\
\indent We found our data often lacked the necessary signal to noise ratio to constrain the reflection parameters. When these parameters were found to be completely unconstrained, they were then fixed to the values recommended by the model documentation. For \texttt{RXTorusD}, the inclination angle was frozen to 45$\degr$ and the ratio of the torus radius was frozen to 0.5\footnote{\url{https://www.astro.unige.ch/reflex/xspec-models}}. \uxc recommended freezing the inner-ring covering factor to 0.4\footnote{\url{https://github.com/JohannesBuchner/xars/blob/master/doc/uxclumpy.rst}}. The user manual has no recommendations regarding inclination angle and cloud dispersion ($\sigma_{tor}$). Therefore, we opted for an intermediate value of 45$\degr$ for the cloud dispersion, and an inclination angle of 45$\degr$ to be consistent with the \rxt value. When either model was unable to constrain the scattering fraction $f_s$, it was frozen to 0.01 \citep[see][]{Ricci2017}. \\
\indent 2MASX J18454978$-$5548252, 2MASX J21570549$+$0632169, and SDSS J073323.77$+$441448.8 were not granted simultaneous XMM observations. Thus, our models for these sources did not include the \texttt{apec} component, as no soft excess below 3\,keV can be detected by \textit{NuSTAR} alone. \\
\indent The details of the fits for each of the 11 sources in our sample are presented in Appendix \ref{app_sec:ind_source}. The best-fit parameters and spectra from each fit can be found in Appendix \ref{app_sec:tab+sepc}. In the following paragraphs, we instead focus on the analysis of the sample as a whole.

\begin{figure*}
    \centering
    \includegraphics[scale=0.6]{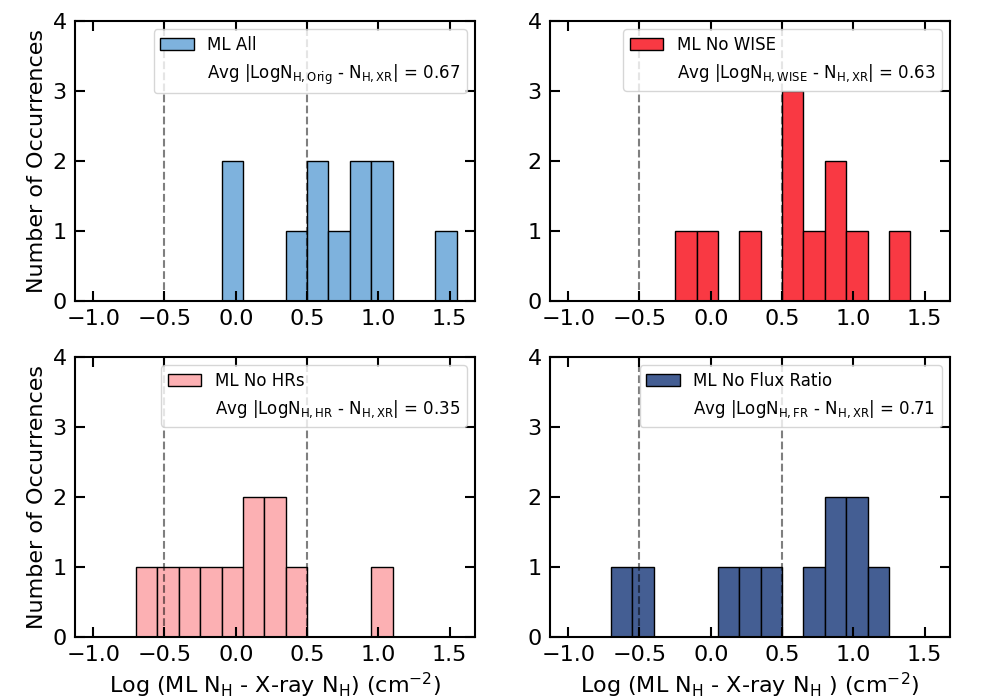}
    \caption{Histograms of the LogN$_{\rm H,los}$ predicted by the machine learning (ML) algorithm subtracted by the \uxc best-fit LogN$_{\rm H,los}$. The top left shows the predictions with all parameters used and the top right shows the predictions without the WISE colors. The bottom left shows the predictions without the \textit{Swift}-XRT hardness ratios and the bottom right shows the predictions without the IR $-$ X-ray flux ratio. The two vertical lines in each plot designate the region where the difference is $< \pm 0.5$. In the legend of each plot, we list the average of the absolute value of LogN$_{\rm H,ML} -$ LogN$_{\rm H,XR}$. }
    \label{fig:ml_param_rem}
\end{figure*}

\begingroup
\renewcommand*{\arraystretch}{2}
\begin{table*}
    \centering
    \caption{\textbf{Notes:} N$_{\rm H}$ comparison between UXClumpy and RXTorusD. N$_{\rm H, los}$ in units of 10$^{23}$ cm$^{-2}$. Horizontal line divides the \textit{NuSTAR} cycles 9 and 10 sources.}
    \label{tab:uxc_vs_rxt}
    \begin{tabular}{lccc}
    \hline \hline
    \textbf{Name}  & \textbf{UXClumpy} & \textbf{RXTorusD}  \\
    \hline
    IGR J19118$-$1707 & 0.47$^{+0.02}_{-0.04}$ & 0.61$^{+0.06}_{-0.05}$ \\
    NGC 4250 & 2.60$^{+0.53}_{-0.43}$ & 9.02$^{+7.28}_{-6.72}$ \\
    MCG--02--34--058 & 7.71$^{+1.65}_{-0.41}$ & 8.75$^{+0.79}_{-1.15}$ \\
    2MASX J18454978$-$5548252 & 1.55$^{+0.29}_{-0.43}$ & 7.12$^{+5.26}_{-3.33}$ \\
    2MASX J21570549$+$0632169 & 0.77$^{+0.15}_{-0.13}$ & 0.76$^{+0.20}_{-0.15}$ \\
    SDSS J073323.77$+$441448.8 & 1.25$^{+0.40}_{-0.32}$ & 1.16$^{+0.35}_{-0.31}$ \\
    \hline
    2MASX J13100789$-$1711398 & 3.58$^{+0.35}_{-0.39}$ & 8.67$^{+2.65}_{-2.14}$ \\
    LEDA 2019751 & 0.52$^{+0.04}_{-0.04}$ & 0.65$^{+0.05}_{-0.03}$ \\
    NGC 52 & 1.17$^{+0.08}_{-0.09}$ & 1.21$^{+0.21}_{-0.11}$ \\
    UGC 12566 & 1.59$^{+0.40}_{-0.37}$ & 2.02$^{+0.59}_{-0.49}$ \\
    NGC 6657 & 1.26$^{+0.06}_{-0.18}$ & 2.25$^{+2.15}_{-0.72}$ \\
    \hline \hline
    \end{tabular}
\end{table*}
\endgroup

%\section{Discussion} \label{sec:disc}

\subsection{Measured NH vs ML-Predicted NH}
This is the first work analyzing the \nus and \xmm data of 11 CT-AGN candidates found by the algorithm described in \citetalias{Silver2023}. Therefore, this section will compare the best-fit X-ray column densities with the predictions of \citetalias{Silver2023} and another method described within, \cite{Asmus2015}. \\
\indent The method of \citetalias{Silver2023} was described above (see Section \ref{sec:intro}). The second method included in this work, \cite{Asmus2015}, uses a simpler framework with only one parameter needed to predict line-of-sight column densities: the ratio between the 12$\mu$m and 2$-$10\,keV fluxes. The authors compiled a sample of 152 AGN with published column densities determined by X-ray spectral modeling. These sources were used to calibrate a relation between the column density and the ratio of the 12$\mu$m and 2$-$10\,keV fluxes (see Figure 23 and Equation 7 in \citealt{Asmus2015}). \\
\indent Figure \ref{fig:xr_vs_ml} shows the \uxc best-fit N$_{\rm H,los}$ values compared with the predictions from the \citetalias{Silver2023} algorithm and the \cite{Asmus2015} relation. We used the Spearman correlation coefficient (SCC) to quantify how correlated each prediction method was with the final X-ray fitting results. The \cite{Asmus2015} relation had an SCC $=$ 0.39, while the \citetalias{Silver2023} algorithm had an SCC $=$ -0.17. This suggests that the \cite{Asmus2015} relation more accurately predicted the N$_{\rm H,los}$ for the 11 sources in this sample, however neither method has a high enough coefficient to claim a statistically significant correlation. \\
\indent It is clear from Figure \ref{fig:xr_vs_ml} that both methods systematically overpredicted the column density of this sample. This could be due to multiple factors. First, we note that AGN have been shown to vary in column density over time \citep[see e.g.,][]{Risaliti2002, Markowitz2014, TorresAlba2023, Pizzetti2025, TorresAlba2025}, which can lead to disagreements between multi-epoch observations. However, we would expect this to result in scatter (i.e. both under and overestimating N$_{\rm H,los}$) rather than a systematic overestimation, which implies other factors must also be at play. \\
\indent We must consider the possibility that certain parameters in our algorithm may be introducing a bias towards higher column densities. To test this, we removed each parameter one at a time and reran the ML algorithm. We then compared the new LogN$_{\rm H,los}$ predictions with the \uxc best-fit LogN$_{\rm H,los}$ values. Figure \ref{fig:ml_param_rem} shows a histogram of the differences for four different versions of the algorithm: all parameters included (top left), WISE colors removed (top right), \textit{Swift}-XRT hardness ratios removed (bottom left), and the IR$-$X-ray flux ratio removed (bottom right)\footnote{These sources did not have \textit{Swift}-BAT data. \citetalias{Silver2023} notes that the BAT data were the least useful to the algorithm and thus the results could still be trusted on sources without them.}. As can be seen in the figure, the algorithm without hardness ratios yielded the most accurate results, with the smallest average of $\lvert$LogN$_{\rm H,ML} -$ LogN$_{\rm H,XR} \rvert$ = 0.35. This makes sense for two reasons: 1) the HRs were the parameter with the most impact on the algorithm's predictions (see Fig. 5 in \citetalias{Silver2023}) and 2) the lowest HR2 (bands 1$-$2\,keV and 2$-$10\,keV) in this sample was 0.77. These two facts together inform us that these sources had large HR values, often indicative of heavy obscuration, and that the algorithm used HRs as its primary method to predict the column density. Therefore, we can say that these 11 sources had high HR values which led the ML algorithm to over-estimate their line-of-sight column densities. This agrees with previous works that have found that solely using HRs to predict N$_{\rm H,los}$ can lead to biased results based on which components were included in the modeling \citep[e.g. the inclusion of a soft excess or reflection component; see][]{Brightman2012, Lambrides2020, Peca2021, Cox2023}. A future work will revise the algorithm from \citetalias{Silver2023} in an attempt to improve its predictive power.

\begin{figure*}[h!]
\begin{center}
\begin{tabular}{ll}
\hspace{-1.4cm}
\includegraphics[scale=0.67]{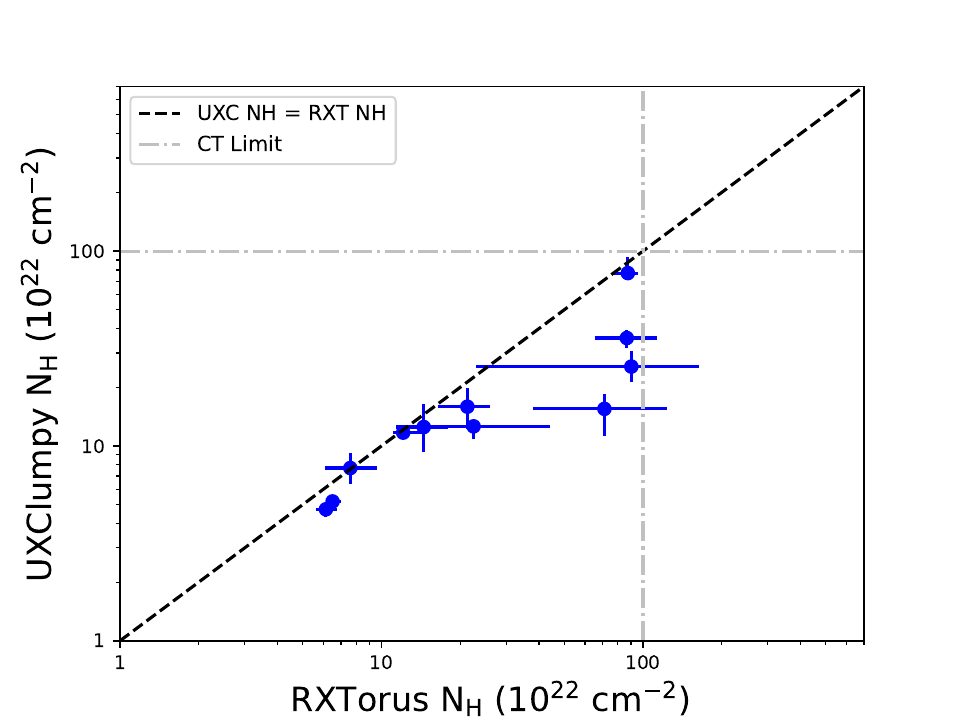}&
\hspace{-1.3cm}
\includegraphics[scale=0.67]{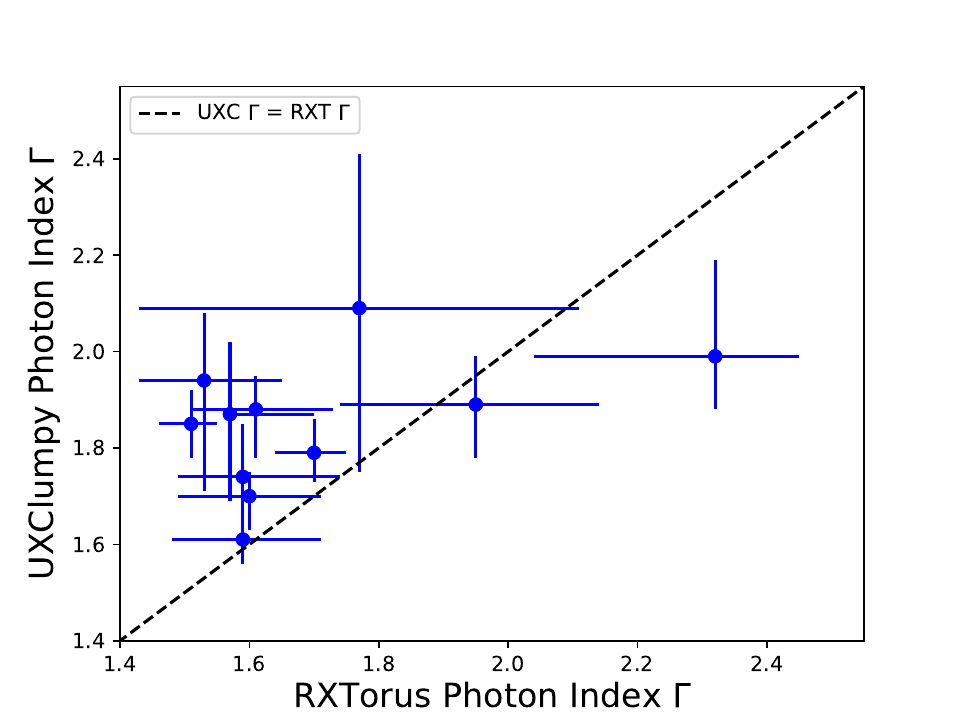}
\end{tabular}
\end{center}
\vspace{-0.6cm}
\caption{
{\textbf{Left:} Scatter plot of the best-fit line-of-sight column densities from \uxc and \texttt{RXTorusD}. The black dotted line represents where N$_{\rm H,los,UXC}$ = N$_{\rm H,los,RXT}$. The grey dash-dotted lines signify the lower boundary of the CT regime. \textbf{Right:} Scatter plot of the best-fit photon indices for all 11 sources from \uxc and \texttt{RXTorusD}. The black dotted line represents where $\Gamma_{\rm UXC}$ = $\Gamma_{\rm RXT}$.}}
\label{fig:uxc_rxt_comps}
\vspace{-0.cm}
\end{figure*}

\subsection{\uxc vs \rxt}
\begin{figure*} 
    \centering
    \includegraphics[scale=0.8]{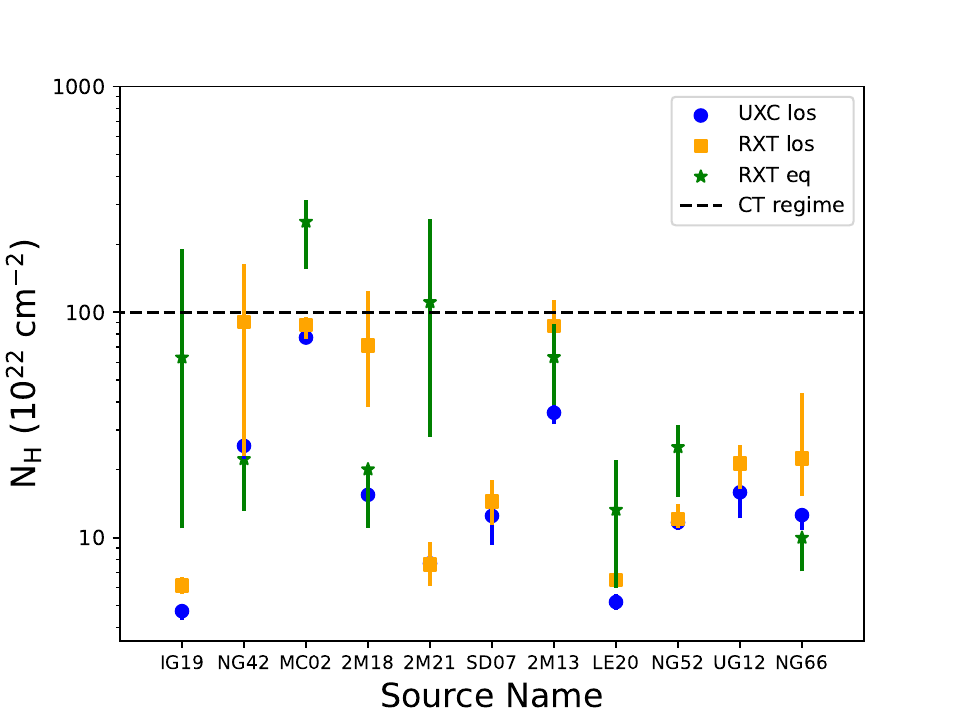}
    \caption{Comparison of column density values between the \texttt{UXClumpy} $N_{H, l.o.s.}$ (blue circles), \texttt{RXTorusD} $N_{H, l.o.s.}$ (orange squares), and \rxt $N_{H, eq.}$ (green stars). If no green star appears, the $N_{H, eq.}$ was fixed to $N_{H, l.o.s.}$ for this source. The dashed horizontal line represents the boundary where any source above it is CT.}
    \label{fig:uxc_vs_rxt_los_eq}
\end{figure*}

\rxt is a new model with limited use thus far \citep[see e.g.,][]{Tortosa2024, Peca2025}, so we wanted to compare its results to a more widely used model, \texttt{UXClumpy}. Table \ref{tab:uxc_vs_rxt} lists the best-fit line-of-sight column densities from both models for all 11 sources. Additionally, Figure \ref{fig:uxc_rxt_comps} visually displays the comparisons between \uxc and \rxt for the best-fit line-of-sight column densities N$_{\rm H,los}$ and the photon indices $\Gamma$. Visually, both models seem to have relatively good agreement. To quantify this, we used Equation 10 from \cite{Cox2023} to determine how many sources had \uxc and \rxt values that agreed with each within the 90\% confidence level. This equation takes the form of:

\begin{equation} \label{eq:90conf_lev}
    \chi^2 = \frac{(NH_{uxc}-\mu)^2}{\delta^2NH_{uxc}} + \frac{(NH_{rxt}-\mu)^2}{\delta^2NH_{rxt}},
\end{equation}

where $\delta$NH is the uncertainty of each N$_{\rm H,los}$ measurement and $\mu$ is the mean of the \uxc and \rxt N$_{\rm H,los}$ for each source. When this value of $\chi^2$ is greater than the critical value of $\chi^2_c$ = 2.706, the two N$_{\rm H,los}$ measurements do not agree within a 90\% confidence level. We found 5/11 agree for N$_{\rm H,los}$ and 8/11 agree for $\Gamma$. The standard deviations from the 1:1 relation line are 24.1 $\times 10^{22}$ cm$^{-2}$ for the N$_{\rm H,los}$ plot and 0.21 for the $\Gamma$ plot. The source by source comparisons are discussed in Appendix \ref{app_sec:ind_source}. \\ %The seven sources that do not agree for N$_{\rm H,los}$ are: NGC 4250, 2MASX J18454978$-$5548252, 2MASX J2157, SDSS J0733, 2MASX J1310, LEDA 2019751, and NGC 52. The three sources that do not agree for $\Gamma$ are: 2MASX J2157, NGC 52 and NGC 6657. 2MASX J2157 and NGC 52 are the only two sources where \rxt and \uxc do not agree on either parameter within the 90\% confidence level. \\
\indent In addition to the general (dis)agreement between the N$_{\rm H,los}$ and $\Gamma$, there exist a few potential trends in these results. First, as can be seen in Figure \ref{fig:uxc_rxt_comps}, \rxt tends to systematically yield higher N$_{\rm H,los}$ values than \texttt{UXClumpy}. %\uxc tends to yield higher N$_{\rm H,los}$ when N$_{\rm H} \lesssim$ 2 $\times$ 10$^{23}$ cm$^{-2}$ and \rxt tends to yield higher N$_{\rm H,los}$ when N$_{\rm H} \gtrsim$ 5 $\times$ 10$^{23}$ cm$^{-2}$. 
Another potential trend regards how \rxt treats the line-of-sight and equatorial column densities. Out of the 9 sources where N$_{\rm H,eq}$ could be constrained, 4 of them (NGC 4250, 2MASX J18454978$-$5548252, 2MASX J13100789$-$1711398, and NGC 6657) had N$_{\rm H,eq}$ values that more closely agreed with the \uxc N$_{\rm H,los}$ than the \rxt N$_{\rm H,los}$ did (see Figure \ref{fig:uxc_vs_rxt_los_eq}). Our team is currently working on another paper (Gianolli et al. in prep.) that analyzes an additional 11 sources with \uxc and \texttt{RXTorusD}. This will double our current sample, thus enabling us to determine if this N$_{\rm H,eq}$ trend has any validity or was just an artifact of small statistics. \\
\indent We note that other recent works have compared \rxt with other models. \cite{Peca2025} analyzed a sample of 21 AGN with \texttt{UXClumpy}, \texttt{RXTorusD}, and \texttt{MYTorus} decoupled. They found the three models yielded consistent results within uncertainties for the line-of-sight column densities  and no significant offsets. Moreover, Gianolli et al. in prep analyzed 11 sources with \texttt{UXClumpy}, \texttt{RXTorusD}, and \texttt{X-skirtor} (Vander Meulen et al. in prep) and found the column densities and photon indices agreed within uncertainties for the majority of the sample.

\section{Summary and Conclusions} \label{sec:summ_+_conc}
The CXB is predicted to be produced by a much larger amount of CT-AGN than have thus far been detected. Working towards resolving this discrepancy, \citetalias{Silver2023} created a machine learning algorithm capable of predicting line-of-sight column density in an accurate and efficient fashion. In this work, we analyzed the first 11 CT-AGN candidates predicted by this method to have a LogN$_{\rm H,los} >$ 23.8. The spectra of these 11 sources were modeled
using the physically motivated models \uxc and \texttt{RXTorusD}. The main results from this work are listed below: 

\begin{itemize}
    \item Based on the best-fit \uxc results, of the 11 AGN, 3 were found to have best-fit line-of-sight column densities 22.7 $<$ LogN$_{\rm H,los} \leq$ 23.0, 5 were found to have 23.0 $<$ LogN$_{\rm H,los} \leq$ 23.25, and 3 were found to have 23.4 $<$ LogN$_{\rm H,los} \leq$ 23.9. According to \texttt{RXTorusD}, 3 sources have uncertainty values that cross into the CT regime. 

    \item This work serves as an additional test \citep[see e.g.,][Gianolli et al. in prep]{Peca2025} of the newly released model \texttt{RXTorusD}, whose results we compared with those obtained using the \uxc model. We find that 8/11 sources have photon indices that agree within the 90\% confidence level and 5/11 sources have line-of-sight column densities that agree within the 90\% confidence level. We note that in the small sample size of nine sources that could constrain the equatorial N$_{\rm H}$ in \texttt{RXTorusD}, four of them had N$_{\rm H,eq}$ that agreed more closely with the N$_{\rm H,los,UXC}$ than the N$_{\rm H,los,RXT}$ did. 

    \item This work is also the first test of candidate CT-AGN found by the new machine learning algorithm in \citetalias{Silver2023}. The predictions of \citetalias{Silver2023} yielded a Spearman correlation coefficient of -0.17, while the predictive relation from \cite{Asmus2015} had a Spearman coefficient of 0.39. While the algorithm from \citetalias{Silver2023} performed worse than the \cite{Asmus2015} relation, it also did not incorrectly predict any unobscured AGN as heavily obscured or CT-AGN. Such false positives were a more frequent occurrence from \cite{Asmus2015} (see Figure 4 in \citetalias{Silver2023}). 

    %\item A future work from our team, Gianolli et al. in prep, will analyze another 11 sources using \uxc and \texttt{RXTorusD}. This will double the sample of AGN analyzed by \texttt{RXTorusD}, allowing us to further test its performance compared with other published models.

    \item Our group is working on an updated machine learning algorithm that will better enable us to narrow the gap between the theoretical CT-AGN fraction and the measured CT-AGN fraction.

\end{itemize}

\section{Acknowledgments}
The material is based upon work supported by NASA under award number 80GSFC24M0006. \\
\indent The authors thank the anonymous referee for their helpful comments which greatly improved our paper. \\
\indent VEG acknowledges funding under NASA contract 80NSSC24K1403.

\appendix

\section{Individual Source Fitting Results} \label{app_sec:ind_source}

\subsection{IGR J19118$-$1707}
Based on the best-fit results listed in Table \ref{tab:igr1918}, both models agree IGR J19118$-$1707 is an obscured AGN, with N$_{\rm H,los} \approx$ 5--6$\times$10$^{22}$ cm$^{-2}$. While the models produced similar values, they do not agree within uncertainties at the 90\% confidence level. %In order to confirm this column density was not inflated by sophisticated models, we performed a test with a simple absorbed power-law and found a consistent N$_{\rm H}$. 
The equatorial N$_{\rm H,eq}$ found by \rxt is heavily obscured, with a value of  6.3$^{+12.7}_{-5.2}$ $\times$ 10$^{23}$ cm$^{-2}$. \\%, suggesting that the torus may reach CT obscuration through more edge-on inclinations. This may have influenced the machine learning algorithm into predicting it was a CT-AGN with N$_{\rm H,los}$ = 3$\times$10$^{24}$ cm$^{-2}$. \\
\indent There is strong agreement between the photon indices, with both models yielding a best-fit $\Gamma \approx$ 1.6. Neither model could constrain the inclination angle, so they were frozen to 45$\degr$. \uxc found a low value for the cloud dispersion with \texttt{TORsigma} $=$ 15$^{+u}_{-2}\degr$. However, \rxt found a high value for the torus radii ratio of r/R $=$ 0.84$^{+u}_{-0.45}$, suggesting a potentially larger torus than that found by \texttt{UXClumpy}.

\subsection{NGC 4250}
The best-fit results for NGC 4250 suggest it is a heavily obscured AGN, however the two models do not agree within the 90\% confidence level. The \uxc fit yields an N$_{\rm H,los}$ = 2.6$^{+0.5}_{-0.4}$ $\times$10$^{23}$ cm$^{-2}$, while the \rxt fit yields an N$_{\rm H,los}$ = 9.0$^{+7.3}_{-6.7}$ $\times$10$^{23}$ cm$^{-2}$. The \rxt upper uncertainty at the 90\% confidence level puts the source into the Compton-thick regime, with a maximum N$_{\rm H,los}$ $\approx$ 1.6$\times$10$^{24}$ cm$^{-2}$.
We note that the \rxt equatorial column density (N$_{\rm H,eq}$ = 2.2$^{+u}_{-0.9}$ $\times$10$^{23}$ cm$^{-2}$) is closer to the \uxc best-fit value. \\
\indent The two models agree with a best-fit photon index around $\Gamma \approx$ 1.7. However, \uxc was unable to constrain any of the reflection parameters nor the scattering fraction. \rxt was unable to constrain the torus radius ratio. However, \rxt was able to partially constrain the inclination angle with a best-fit value of 74$^{+u}_{-70}$$\degr$ (see Table \ref{tab:ngc4250}).

\subsection{MCG--02--34--058}
MCG--02--34--058 was found to be the most obscured source in our sample (using the average of the two model results), with \uxc finding a best-fit N$_{\rm H,los}$ = 7.7$^{+1.7}_{-0.4}$ $\times$10$^{23}$ cm$^{-2}$ and \rxt yielding a best-fit N$_{\rm H,los}$ = 8.9$^{+0.8}_{-1.2}$ $\times$10$^{23}$ cm$^{-2}$. Additionally, the equatorial column density is largest in our sample (N$_{\rm H,eq}$ = 25.1$^{+6.2}_{-9.6}$ $\times$10$^{23}$ cm$^{-2}$). \\
\indent This source also displayed relatively different photon index values (1.99$^{+0.20}_{-0.11}$ vs 2.32$^{+0.13}_{-0.28}$), although they agree within the 90\% confidence level (see Table \ref{tab:mcg02}). Both models suggest the torus covers a significant amount of the volume surrounding the black hole, with the best fits yielding very high values for \texttt{CTKcover} and r/R (0.60$^{+u}_{-0.04}$ and 0.98$^{+u}_{-0.04}$), and a moderately large value for \texttt{TORsigma} (40.9$^{+33.1}_{-32.4}$).

\subsection{2MASX J18454978$-$5548252}
The best-fit results for 2MASX J18454978$-$5548252 suggest it is a heavily obscured AGN, however the models do not agree within the 90\% confidence level. \uxc yielded a column density of 1.6$^{+0.3}_{-0.4}$ $\times$10$^{23}$ cm$^{-2}$ and \rxt yielded N$_{\rm H,los}$ = 7.1$^{+5.3}_{-3.3}$ $\times$10$^{23}$ cm$^{-2}$. The equatorial column density, N$_{\rm H,eq}$ = 2.0$^{+u}_{-0.9}$ $\times$10$^{23}$ cm$^{-2}$, agrees with the \uxc N$_{\rm H,los}$ value. \\
\indent The photon indices show good agreement, yielding values of $\Gamma \approx$ 1.65. Neither \texttt{CTKcover} nor the scattering fraction could be constrained and thus they were frozen to our pre-determined values. Similarly to MCG--02--34--058, the values found for \texttt{TORsigma} and r/R (84$^{+u}_{-44}$ and 1.0$^{+u}_{-0.43}$) suggest the torus is very spread out (see Table \ref{tab:2m1845}).

\subsection{2MASX J21570549$+$0632169}
Both models agree 2MASX J21570549$+$0632169 is an obscured AGN within the 90\% confidence level. \uxc found an N$_{\rm H,los}$ = 7.7$^{+1.5}_{-1.3}$ $\times$10$^{22}$ cm$^{-2}$ and \rxt found an N$_{\rm H,los}$ = 7.6$^{+0.2}_{-0.2}$ $\times$10$^{22}$ cm$^{-2}$. Despite having one of the lowest N$_{\rm H,los}$ values in our sample, the equatorial column density was the second highest and in the CT regime, with a value of N$_{\rm H,eq}$ = 1.1$^{+1.5}_{-0.8}$ $\times$10$^{24}$ cm$^{-2}$. \\
\indent The photon indices agree within the 90\% confidence level (1.70$^{+0.05}_{-0.07}$ from \uxc and 1.60$^{+0.11}_{-0.11}$ from \texttt{RXTorusD}). The only reflection parameter that could be constrained was the \uxc inclination angle, which suggests an edge-on view of the torus (90$\degr$ $^{+u}_{-20}$, see Table \ref{tab:2m2157}).

\subsection{SDSS J073323.77$+$441448.8}
The two models agree that SDSS J073323.77$+$441448.8 is a heavily obscured AGN at the 90\% confidence level. \uxc found an N$_{\rm H}$ = 1.3$^{+0.4}_{-0.3}$ $\times$10$^{23}$ cm$^{-2}$ and \rxt found an N$_{\rm H}$ = 1.5$^{+0.4}_{-0.3}$ $\times$10$^{23}$ cm$^{-2}$. We were unable to constrain the equatorial N$_{\rm H}$, thus it was fixed to the N$_{\rm H,los}$ value. \\
\indent The best-fit photon index values do not agree within the 90\% confidence level (1.94$^{+0.14}_{-0.23}$ vs 1.53$^{+0.12}_{-0.10}$). Due to the lack of XMM data, the only reflection parameter that could be constrained was \texttt{CTKcover} = 0.6$^{+u}_{-0.13}$, as seen in Table \ref{tab:sdss07}. The scattering fraction was unable to be constrained as well.

\subsection{2MASX J13100789$-$1711398}
The two line-of-sight column density values of 2MASX J13100789$-$1711398 did not agree within the 90\% confidence level, however, both models found this source to be heavily obscured. \uxc yields an N$_{\rm H}$ = 3.6$^{+0.4}_{-0.4}$ $\times$10$^{23}$ cm$^{-2}$, while \rxt yields a larger value of N$_{\rm H}$ = 8.7$^{+2.7}_{-2.1}$ $\times$10$^{23}$ cm$^{-2}$. 2MASX J13100789$-$1711398 is one of three sources where the upper uncertainty from the \rxt best fit extends into the CT regime. The N$_{\rm H,eq}$ was in between the two line-of-sight values, with N$_{\rm H,eq}$ = 6.3$^{+2.6}_{-2.4}$ $\times$10$^{23}$ cm$^{-2}$. \\ %We elected to model this source using the decoupled configuration of \texttt{NYTorus}\footnote{Since 2MASX J13100789$-$1711398 was the only source in our sample with such a large disagreement in N$_{\rm H,los}$ between \uxc and \texttt{RXTorusD}, we did not feel it necessary to use \myt for any of the other sources.} \citep{Murphy2009}, which as stated above, has a similar geometry to \texttt{RXTorusD}. Using \texttt{MYTorus}, we found a line-of-sight column density of N$_{\rm H,los}$ = 3.3$^{+0.5}_{-0.4}$ $\times$10$^{23}$ cm$^{-2}$ and an average torus column density of N$_{\rm H,avg}$ = 1.0$^{+0.5}_{-0.5}$ $\times$10$^{24}$ cm$^{-2}$. Based on these results, we believe 2MASX J13100789$-$1711398 is a heavily obscured AGN with a CT average torus column density. As found with other sources, \rxt is not always able to disentangle the line-of-sight column density from the average torus column density, which is a potential cause of the observed disagreements. \\
\indent The two models did agree on the photon index, with an average value around $\Gamma \approx$ 1.9. The \texttt{CTKcover} was found to have low a value of 0$^{+0.14}_{-u}$, while both the r/R and \texttt{TORsigma} were best-fit near their maximum values, 0.97$^{+0.02}_{-0.02}$ and 84$^{+u}_{-6}$, respectively. Only \rxt could constrain the inclination angle and it suggests a near face-on viewing of the torus (15$^{+5}_{-3}$). The full fitting results can be seen in Table \ref{tab:2m13}. \\

\subsection{LEDA 2019751}
Both models found that LEDA 2019751 is an obscured AGN with N$_{\rm H}$ $\approx$ 5-6$\times$10$^{22}$ cm$^{-2}$, however they did not agree within the 90\% confidence level. The equatorial column density was found to be heavily obscured, with an N$_{\rm H,eq}$ = 1.3$^{+0.9}_{-0.7}$ $\times$10$^{23}$ cm$^{-2}$. \\
\indent The photon indices agreed within 90\% with an average value around $\sim$1.75. \uxc suggests the torus is compact in size (\texttt{CTKcover} = 0$^{+
0.15}_{-u}$ and \texttt{TORsigma} = 11$^{+41}_{-7}$) and is being viewed edge-on (90$^{+u}_{-44} \degr$). However, \rxt disagrees with both statements, suggesting the torus is large in size (r/R = 1.0$^{+u}_{-0.88}$) and we are viewing it nearly face-on (3$^{+13}_{-u} \degr$). Neither scattering fraction could be constrained (see Table \ref{tab:leda20}).

\subsection{NGC 52}
NGC 52 was found to be heavily obscured in both models, which agree within the 90\% confidence level. \uxc found a best-fit N$_{\rm H}$ $=$ 1.2$^{+0.1}_{-0.1}$ $\times$10$^{23}$ cm$^{-2}$ and \rxt found a value of N$_{\rm H}$ = 1.2$^{+0.2}_{-0.1}$ $\times$10$^{23}$ cm$^{-2}$. The equatorial column density was found to be the largest of the three, with a value of N$_{\rm H,eq}$ = 2.5$^{+0.6}_{-1.0}$ $\times$10$^{23}$ cm$^{-2}$. \\
\indent NGC 52 was one of three sources in our sample where the photon index from each model did not agree within the 90\% confidence level. \rxt had a significantly harder photon index than \uxc (1.51$^{+0.04}_{-0.05}$ vs 1.85$^{+0.07}_{-0.07}$, see Table \ref{tab:ngc52}). The torus appears to be rather spread out, with a \tor = 42$^{+31}_{-16}$ and an r/R = 1.0$^{+u}_{-0.10}$. However, the inner ring was found to have the lowest possible covering factor, \ctk = 0.0$^{+0.3}_{-u}$. Both models suggest we are viewing the torus near face-on, with inclination angles of 0$^{+68}_{-u} \degr$ and 26$^{+23}_{-6} \degr$ from \uxc and \texttt{RXTorusD}, respectively. Neither model could constrain the scattering fraction nor the temperature of the soft excess.

\subsection{UGC 12566}
UGC 12566 had the shortest \nus observation ($\sim$21\,ks) and a lower count rate than expected from the cycle 9 proposal, and thus had the fewest counts in our sample. Consequently, none of the reflection parameters could be constrained. Both models show good agreement with an N$_{\rm H}$ classification of heavily obscured (N$_{\rm H}$ $\approx$ 1.6---2.1$\times$10$^{23}$ cm$^{-2}$) and a relatively soft photon index of $\Gamma \approx$ 1.8--2.1. The N$_{\rm H,eq}$ had to be fixed to the N$_{\rm H,los}$ in order to achieve a good fit (see Table \ref{tab:ugc12}). The constant controlling the scattering fraction f$_s$ was frozen to 0.01 in the \rxt fit.

\subsection{NGC 6657}
Both fits suggest NGC 6657 is a heavily obscured AGN, but they do not agree within the 90\% confidence level. \uxc yields a best-fit column density of N$_{\rm H,los}$ $=$ 1.3$^{+0.1}_{-0.2}$ $\times$10$^{23}$ cm$^{-2}$, while \rxt yields a value of N$_{\rm H,los}$ $=$ 2.3$^{+2.2}_{-0.7}$ $\times$10$^{23}$ cm$^{-2}$. The equatorial column density is the lowest of the three, with a value of N$_{\rm H,eq}$ = 1.0$^{+u}_{-0.3}$ $\times$10$^{23}$. \\
\indent NGC 6657 is the third source in the sample where the photon indices do not agree within the 90\% confidence level ($\Gamma_{\rm UXC}$=1.88$^{+0.07}_{-0.10}$ vs $\Gamma_{\rm RXT}$=1.61$^{+0.12}_{-0.10}$). There was no soft excess found for this source. Therefore, the normalization of the \texttt{apec} component was frozen to 0 in both models. \uxc found relatively small values regarding the size of the torus (\ctk = 0.2$^{+0.2}_{-0.1}$ and \tor = 21$^{+u}_{-3}$), while the torus radius ratio in \rxt suggests a nearly circular torus (r/R = 0.95$^{+u}_{-0.34}$). The inclination angle from \uxc had to be fixed to 45$\degr$, while the \rxt inclination angle was fit to 50$^{+u}_{-19} \degr$. The \uxc scattering fraction could not be constrained and thus was frozen to 0.01.% (see Table \ref{tab:ngc6657}).

\section{Best-Fit Parameters and Spectra} \label{app_sec:tab+sepc}

In the tables below, we present the best-fit results for each source from the \uxc and \rxt fits. The parameters obtained are as follows: \textit{$\chi^2$/dof} is the $\chi^2$ divided by the degrees of freedom, \textit{kT} is the temperature of the soft excess determined by \texttt{apec} in units of keV, and $\Gamma$ is the power law photon index. \textit{norm} is the main power-law normalization (in units of photons cm$^2$ s$^{-1}$ keV$^{-1}$ $\times$ 10$^{-2}$) measured at 1 keV, \textit{CTKcover} is the covering factor of the inner ring of clouds as computed with \texttt{UXClumpy}, and \textit{Tor $\sigma$} is the cloud dispersion factor as computed with \texttt{UXClumpy}. \textit{$\theta_{obs}$} is the inclination angle and \textit{r / R} is the inner-to-outer radius ratio of the torus. \textit{$N_{H, l.o.s.}$} is the line-of-sight hydrogen column density in units of 10$^{23}$ cm$^{-2}$ and \textit{$N_{H, e.q.}$} is the equatorial torus hydrogen column density in units of 10$^{23}$ cm$^{-2}$. $f_s$ is the constant controlling the scattered continuum and $c_{nus}$ is the cross-calibration constant between the \xmm and \nus data. \\
\indent F$\rm_{2-10\,keV}$ and F$\rm_{15-55\,keV}$ are the observed fluxes in the 2$-$10 and 15$-$55\,keV bands in units of erg cm$^{-2}$ s$^{-1}$. L$\rm_{2-10\,keV}$ and L$\rm_{15-55\,keV}$ are the intrinsic luminosities in the 2$-$10 and 15$-$55\,keV bands in units of erg s$^{-1}$. ``*'' indicates the parameter was frozen to that value during fitting. ``u'' listed in the errors designates the parameter was unconstrained in that direction. \\
\indent In all the spectra below, the XMM MOS1 data is plotted in black, the MOS2 data in red, and the PN data in green. The \nus FPMA data is in blue and the FPMB data is in cyan. The solid lines represent the line-of-sight continuum, the dashed lines represent the reflection continuum, and the dotted lines represent the soft excess component.

\begingroup
\renewcommand*{\arraystretch}{2}
\begin{table*}
    \centering
    \label{tab:igr1918}
    \begin{tabular}{lccc}
    \textbf{IGR J19118$-$1707} \\
    \hline \hline
    \textbf{Model} & \textbf{UXClumpy} & \textbf{RXTorusD} \\
    \hline
    $\chi^2$/dof & 387/358 & 373/356 \\
    kT & 0.69$^{+0.13}_{-0.13}$ & 0.68$^{+0.14}_{-0.12}$ \\
    $\Gamma$ & 1.61$^{+0.08}_{-0.05}$ & 1.59$^{+0.12}_{-0.11}$ \\
    norm $10^{-2}$ & 0.06$^{+0.02}_{-0.01}$ & 0.04$^{+0.01}_{-0.01}$ \\
    CTKcover  & 0.4* & ... \\
    Tor $\sigma$ & 15.3$^{+u}_{-2.0}$ & ... \\
    $\theta_{obs}$ & 45* & 45* \\
    r / R & ... & 0.84$^{+u}_{-0.45}$ \\
    $N_{H, l.o.s.}$ & 0.47$^{+0.02}_{-0.04}$ & 0.61$^{+0.06}_{-0.05}$ \\
    $N_{H, eq.}$ & ... & 6.29$^{+12.70}_{-5.19}$ \\
    $f_s$ 10$^{-2}$ & 4.43$^{+1.90}_{-u}$ & 2.20$^{+0.60}_{-0.70}$ \\
    c$_{nus}$ & 1.20$^{+0.05}_{-0.06}$ & 1.19$^{+0.05}_{-0.06}$ \\
    \hline
    F$\rm_{2-10\,keV}$ & 1.41$^{+0.03}_{-0.03}$ $\times$ 10$^{-12}$ & 1.45$^{+0.03}_{-0.03}$ $\times$ 10$^{-12}$ \\
    F$\rm_{15-55\,keV}$ & 1.32$^{+0.03}_{-0.06}$ $\times$ 10$^{-13}$ & 1.35$^{+0.06}_{-0.03}$ $\times$ 10$^{-13}$ \\
    L$\rm_{2-10\,keV}$ & 3.01$^{+0.75}_{-0.66}$ $\times$ 10$^{42}$ & 2.39$^{+1.38}_{-0.37}$ $\times$ 10$^{42}$ \\
    L$\rm_{15-55\,keV}$ & 7.06$^{+1.77}_{-1.54}$ $\times$ 10$^{42}$ & 5.21$^{+4.00}_{-1.07}$ $\times$ 10$^{42}$ \\
    \hline \hline
    %An absorbed powerlaw yielded an NH$_{los}$ of 0.06.
    \end{tabular}
\end{table*}
\endgroup

\begin{figure*}
    \centering
    \includegraphics[angle=270, scale=0.6]{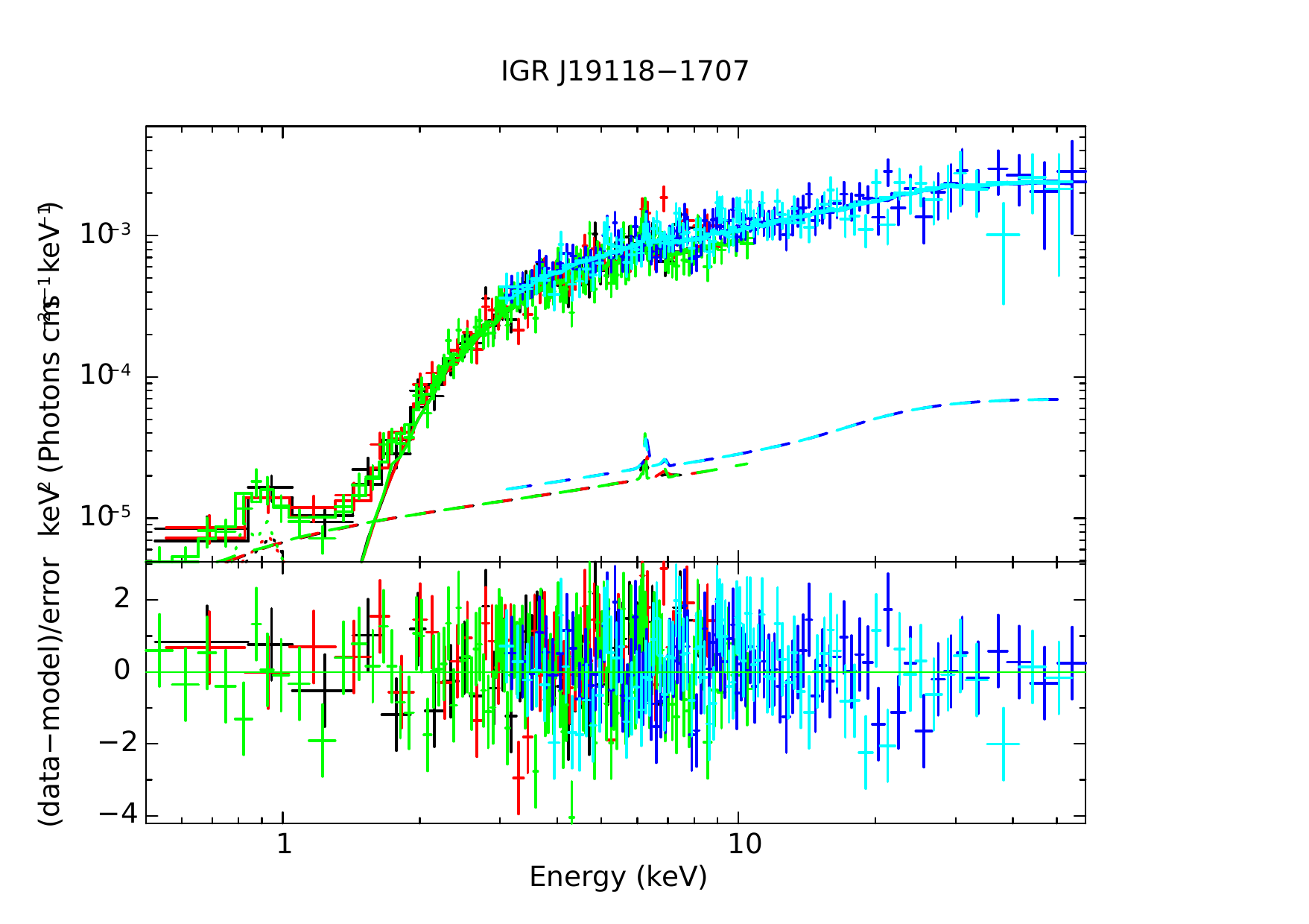}
    \caption{\uxc best fit of IGR J19118$-$1707.}
    \label{fig:igr}
\end{figure*}

\begingroup
\renewcommand*{\arraystretch}{2}
\begin{table*}
    \centering
    \label{tab:ngc4250}
    \begin{tabular}{lccc}
    \textbf{NGC 4250} \\
    \hline \hline
    \textbf{Model} & \textbf{UXClumpy} & \textbf{RXTorusD} \\
    \hline
     $\chi^2$/dof & 122/106 & 106/104 \\
    kT & 0.31$^{+0.04}_{-0.04}$ & 0.30$^{+0.04}_{-0.04}$ \\
    $\Gamma$ & 1.87$^{+0.15}_{-0.18}$ & 1.57$^{+0.13}_{-u}$ \\
    norm $10^{-2}$ & 0.08$^{+0.04}_{-0.05}$ & 0.04$^{+0.02}_{-0.03}$ \\
    CTKcover & 0.4* & ... \\
    Tor $\sigma$ & 45* & ... \\
    $\theta_{obs}$ & 45* & 74.4$^{+u}_{-70}$ \\
    r / R & ... & 0.5* \\
    $N_{H, l.o.s.}$ & 2.60$^{+0.53}_{-0.43}$ & 9.02$^{+7.28}_{-6.72}$ \\
    $N_{H, eq.}$ & ... & 2.23$^{+u}_{-0.91}$ \\
    $f_s$ 10$^{-2}$ & 1* & 0.78$^{+1.80}_{-0.50}$ \\
    c$_{nus}$ & 1.10$^{+0.12}_{-0.11}$  & 1.11$^{+0.12}_{-0.12}$ \\
    \hline
    F$\rm_{2-10\,keV}$ & 5.62$^{+0.27}_{-0.25}$ $\times$ 10$^{-13}$ & 5.50$^{+0.26}_{-0.25}$ $\times$ 10$^{-13}$  \\
    F$\rm_{15-55\,keV}$ & 8.71$^{+0.62}_{-0.77}$ $\times$ 10$^{-14}$ & 9.12$^{+0.65}_{-0.61}$ $\times$ 10$^{-14}$ \\
    L$\rm_{2-10\,keV}$ & 2.02$^{+1.25}_{-1.32}$ $\times$ 10$^{41}$ & 3.49$^{+1.97}_{-3.27}$ $\times$ 10$^{41}$ \\
    L$\rm_{15-55\,keV}$ & 2.98$^{+1.67}_{-1.76}$ $\times$ 10$^{41}$ & 4.67$^{+1.07}_{-1.77}$ $\times$ 10$^{41}$ \\
    \hline \hline
    \end{tabular}
\end{table*}
\endgroup

\begin{figure*}
    \centering
    \includegraphics[angle=270, scale=0.6]{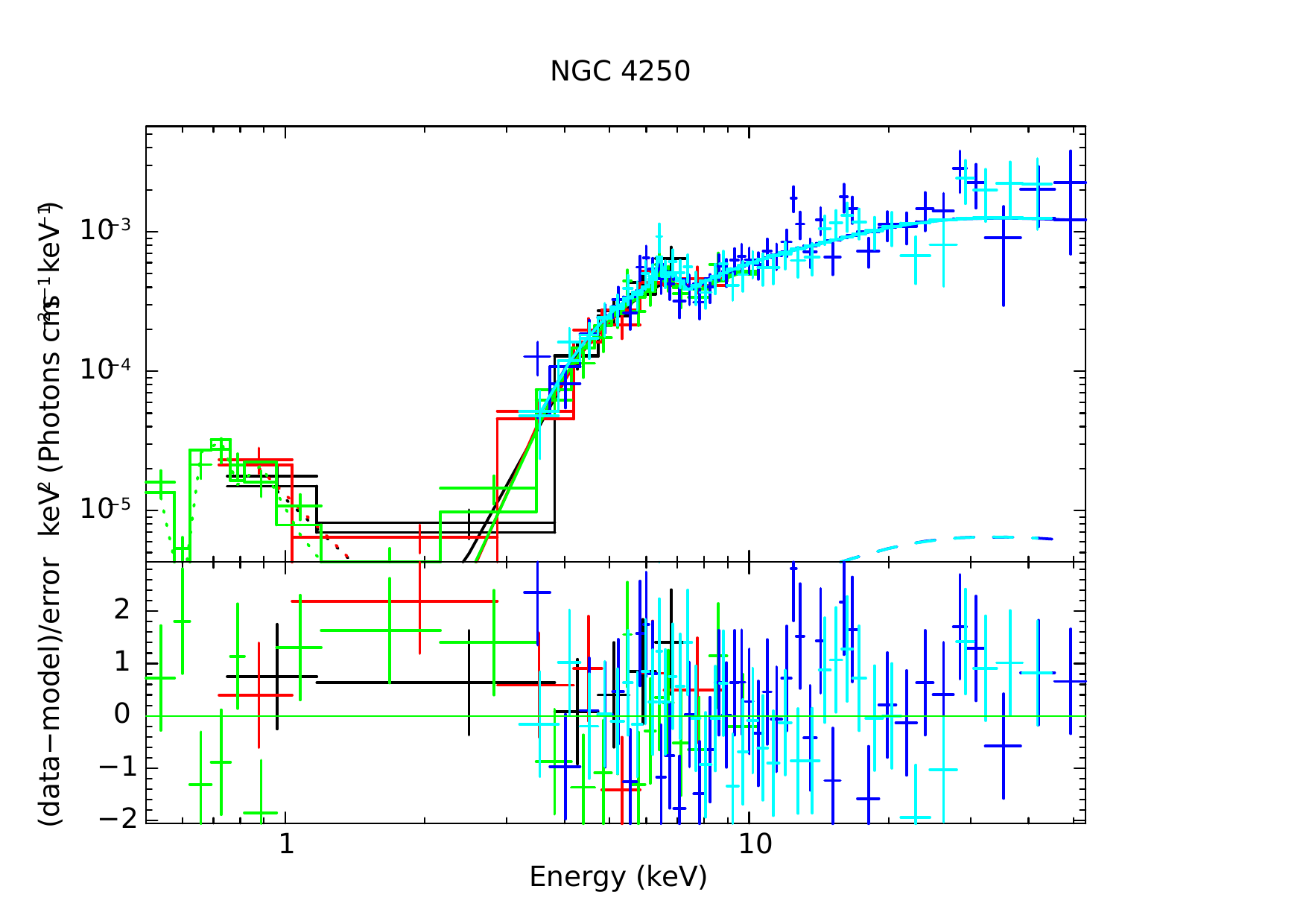}
    \caption{\uxc best fit of NGC 4250.}
    \label{fig:ngc}
\end{figure*}

\begingroup
\renewcommand*{\arraystretch}{2}
\begin{table*}
    \centering
    \label{tab:mcg02}
    \begin{tabular}{lccc}
    \textbf{MCG--02--34--058} \\
    \hline \hline
    \textbf{Model} & \textbf{UXClumpy} & \textbf{RXTorusD} \\
    \hline
     $\chi^2$/dof & 231/197 & 229/197 \\
    kT & 0.29$^{+0.35}_{-0.12}$ & 0.24$^{+u}_{-u}$ \\
    $\Gamma$ & 1.99$^{+0.20}_{-0.11}$ & 2.32$^{+0.13}_{-0.28}$\\
    norm $10^{-2}$ & 0.66$^{+0.74}_{-0.17}$ & 0.99$^{+0.54}_{-0.57}$ \\
    CTKcover & 0.60$^{+u}_{-0.04}$ & ... \\
    Tor $\sigma$ & 40.9$^{+33.1}_{-32.4}$ & ... \\
    $\theta_{obs}$ & 61.1$^{+u}_{-21.4}$ & 33.0$^{+10.2}_{-11.5}$ \\
    r / R & ... & 0.98$^{+u}_{-0.04}$ \\
    $N_{H, l.o.s.}$ & 7.71$^{+1.65}_{-0.41}$ & 8.88$^{+0.79}_{-1.15}$ \\
    $N_{H, eq.}$ & ... & 25.1$^{+6.16}_{-9.64}$ \\
    $f_s$ 10$^{-2}$ & 0.36$^{+0.74}_{-0.19}$ & 0.04$^{+0.06}_{-0.02}$ \\
    c$_{nus}$ & 1.19$^{+0.07}_{-0.08}$ & 1.18$^{+0.08}_{-0.07}$ \\
    \hline
    F$\rm_{2-10\,keV}$ & 9.77$^{+0.46}_{-0.22}$ $\times$ 10$^{-13}$ & 9.82$^{+0.23}_{-0.44}$ $\times$ 10$^{-13}$ \\
    F$\rm_{15-55\,keV}$ & 3.24$^{+0.15}_{-0.15}$ $\times$ 10$^{-13}$ & 3.47$^{+0.16}_{-0.23}$ $\times$ 10$^{-13}$ \\
    L$\rm_{2-10\,keV}$ & 1.44$^{+1.61}_{-0.37}$ $\times$ 10$^{43}$ & 1.70$^{+0.48}_{-0.52}$ $\times$ 10$^{43}$ \\
    L$\rm_{15-55\,keV}$ & 1.67$^{+1.86}_{-0.43}$ $\times$ 10$^{43}$ & 1.35$^{+0.51}_{-0.54}$ $\times$ 10$^{43}$ \\
    \hline \hline
    \end{tabular}
\end{table*}
\endgroup

\begin{figure*}
    \centering
    \includegraphics[angle=270, scale=0.6]{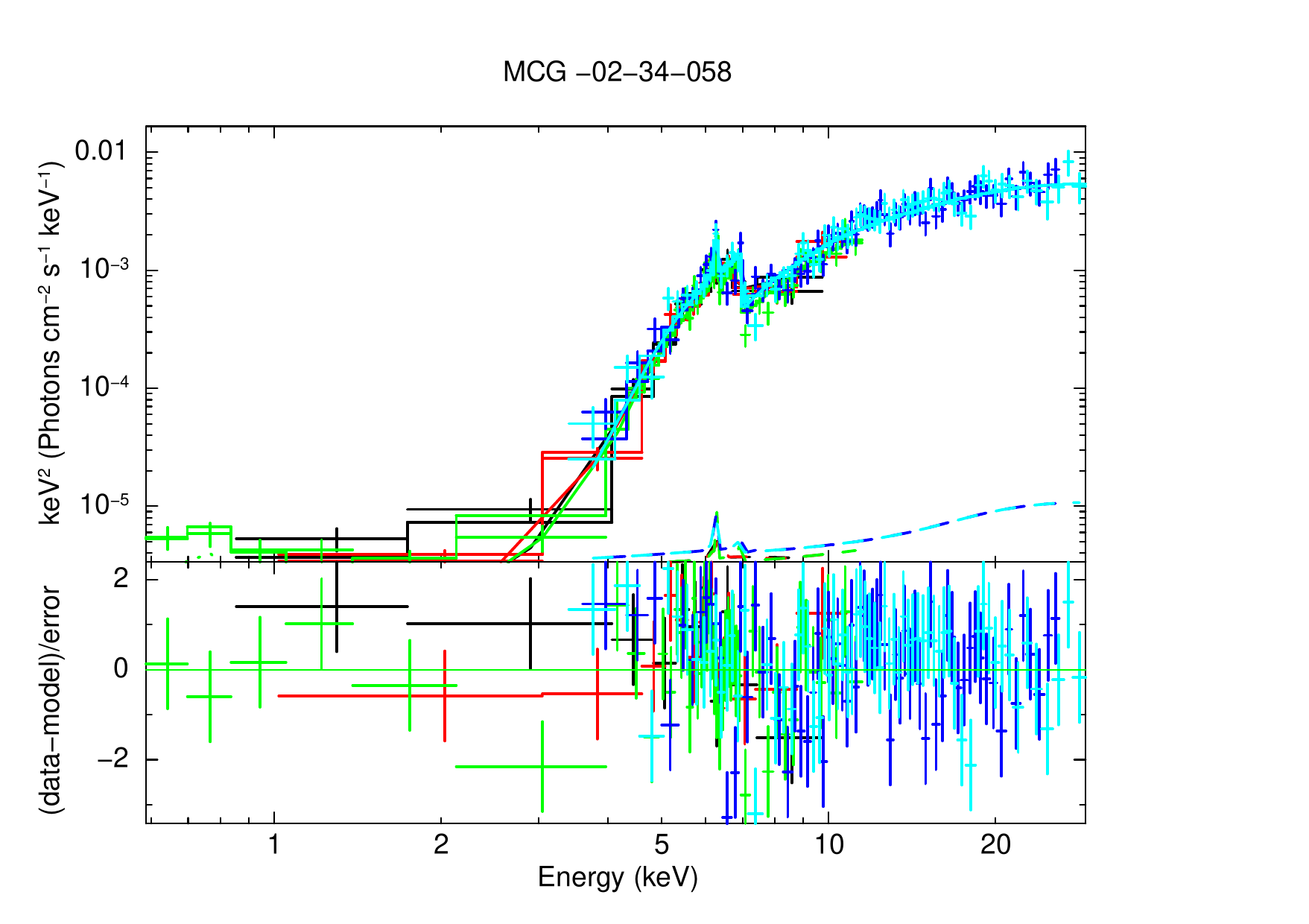}
    \caption{\uxc best fit of MCG--02--34--058.}
    \label{fig:mgc}
\end{figure*}

\begingroup
\renewcommand*{\arraystretch}{2}
\begin{table*}
    \centering
    \label{tab:2m1845}
    \begin{tabular}{lccc}
    \textbf{2MASX J18454978$-$5548252} \\
    \hline \hline
    \textbf{Model} & \textbf{UXClumpy} & \textbf{RXTorusD} \\
    \hline
     $\chi^2$/dof & 115/118  & 110/117 \\
    kT & ... & ... \\
    $\Gamma$ & 1.74$^{+0.11}_{-0.06}$ & 1.59$^{+0.15}_{-0.10}$ \\
    norm $10^{-2}$ & 0.11$^{+0.04}_{-0.02}$ & 0.09$^{+0.07}_{-0.03}$ \\
    CTKcover & 0.4* & ... \\
    Tor $\sigma$ & 84$^{+u}_{-44}$ & ... \\
    $\theta_{obs}$ & 0.0$^{+2.7}_{-u}$ & 23.1$^{+u}_{-12.5}$ \\
    r / R & ... & 1.00$^{+u}_{-0.43}$ \\
    $N_{H, l.o.s.}$ & 1.55$^{+0.29}_{-0.43}$ & 7.12$^{+5.26}_{-3.33}$ \\
    $N_{H, eq.}$ & ... & 2.01$^{+u}_{-0.90}$ \\
    $f_s$ 10$^{-2}$ & 1* & 1* \\
    c$_{nus}$ & ... & ... \\
    \hline
    F$\rm_{2-10\,keV}$ & 1.38$^{+0.07}_{-0.03}$ $\times$ 10$^{-12}$ & 1.41$^{+0.06}_{-0.04}$ $\times$ 10$^{-12}$ \\
    F$\rm_{15-55\,keV}$ & 5.50$^{+0.26}_{-0.13}$ $\times$ 10$^{-12}$ & 5.37$^{+0.25}_{-0.12}$ $\times$ 10$^{-12}$ \\
    L$\rm_{2-10\,keV}$ & 8.90$^{+3.05}_{-1.73}$ $\times$ 10$^{42}$ & 1.45$^{+1.31}_{-0.76}$ $\times$ 10$^{43}$ \\
    L$\rm_{15-55\,keV}$ & 1.43$^{+0.49}_{-0.28}$ $\times$ 10$^{43}$ & 2.27$^{+0.52}_{-0.31}$ $\times$ 10$^{43}$ \\
    \hline \hline
    \end{tabular}
\end{table*}
\endgroup

\begin{figure*}
    \centering
    \includegraphics[angle=270, scale=0.6]{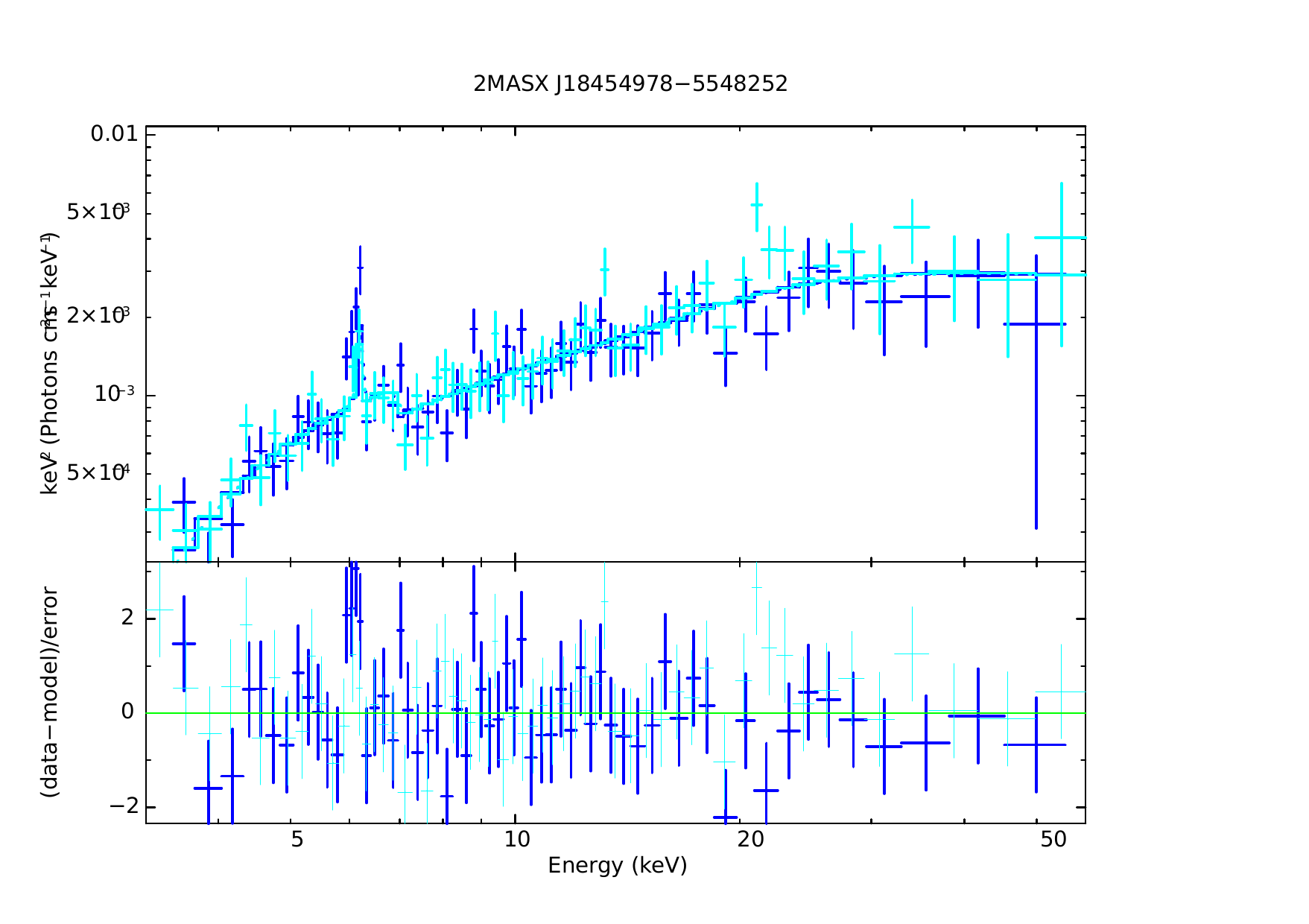}
    \caption{\uxc best fit of 2MASX J18454978$-$5548252.}
    \label{fig:2m18}
\end{figure*}

\begingroup
\renewcommand*{\arraystretch}{2}
\begin{table*}
    \centering
    %\caption{\textbf{Notes:} Same as \ref{tab:igr1918}. An absorbed powerlaw yielded an NH$_{los}$ of 0.8$\times$10$^{23}$ cm$^{-2}$.}
    \label{tab:2m2157}
    \begin{tabular}{lccc}
    \textbf{2MASX J21570549$+$0632169} \\
    \hline \hline
    \textbf{Model} & \textbf{UXClumpy} & \textbf{RXTorusD} \\
    \hline
     $\chi^2$/dof & 276/272 & 279/280 \\
    kT & ... & ... \\
    $\Gamma$ & 1.70$^{+0.05}_{-0.07}$ & 1.60$^{+0.11}_{-0.11}$ \\
    norm $10^{-2}$ & 0.26$^{+0.03}_{-0.03}$ & 0.15$^{+0.05}_{-0.03}$ \\
    CTKcover & 0.4* & ... \\
    Tor $\sigma$ & 45* & ... \\
    $\theta_{obs}$ & 90$^{+u}_{-20}$ & 45* \\
    r / R & ... & 0.5* \\
    $N_{H, l.o.s.}$ & 0.77$^{+0.15}_{-0.13}$ & 0.76$^{+0.02}_{-0.02}$ \\
    $N_{H, eq.}$ & ... & 11.1$^{+14.7}_{-8.3}$ \\
    $f_s$ 10$^{-2}$ & 1* & 1* \\
    c$_{nus}$ & ... & ... \\
    \hline
    F$\rm_{2-10\,keV}$ & 4.79$^{+0.11}_{-0.11}$ $\times$ 10$^{-12}$ & 4.83$^{+0.10}_{-0.11}$ $\times$ 10$^{-12}$ \\
    F$\rm_{15-55\,keV}$ & 1.26$^{+0.03}_{-0.03}$ $\times$ 10$^{-11}$ & 1.24$^{+0.03}_{-0.02}$ $\times$ 10$^{-11}$\\
    L$\rm_{2-10\,keV}$ & 1.75$^{+0.19}_{-0.21}$ $\times$ 10$^{43}$ & 1.55$^{+0.17}_{-1.75}$ $\times$ 10$^{43}$\\
    L$\rm_{15-55\,keV}$ & 2.59$^{+0.28}_{-0.31}$ $\times$ 10$^{43}$ & 2.77$^{+0.43}_{-0.47}$ $\times$ 10$^{43}$\\
    \hline \hline
    \end{tabular}
\end{table*}
\endgroup

\begin{figure*}
    \centering
    \includegraphics[angle=270, scale=0.6]{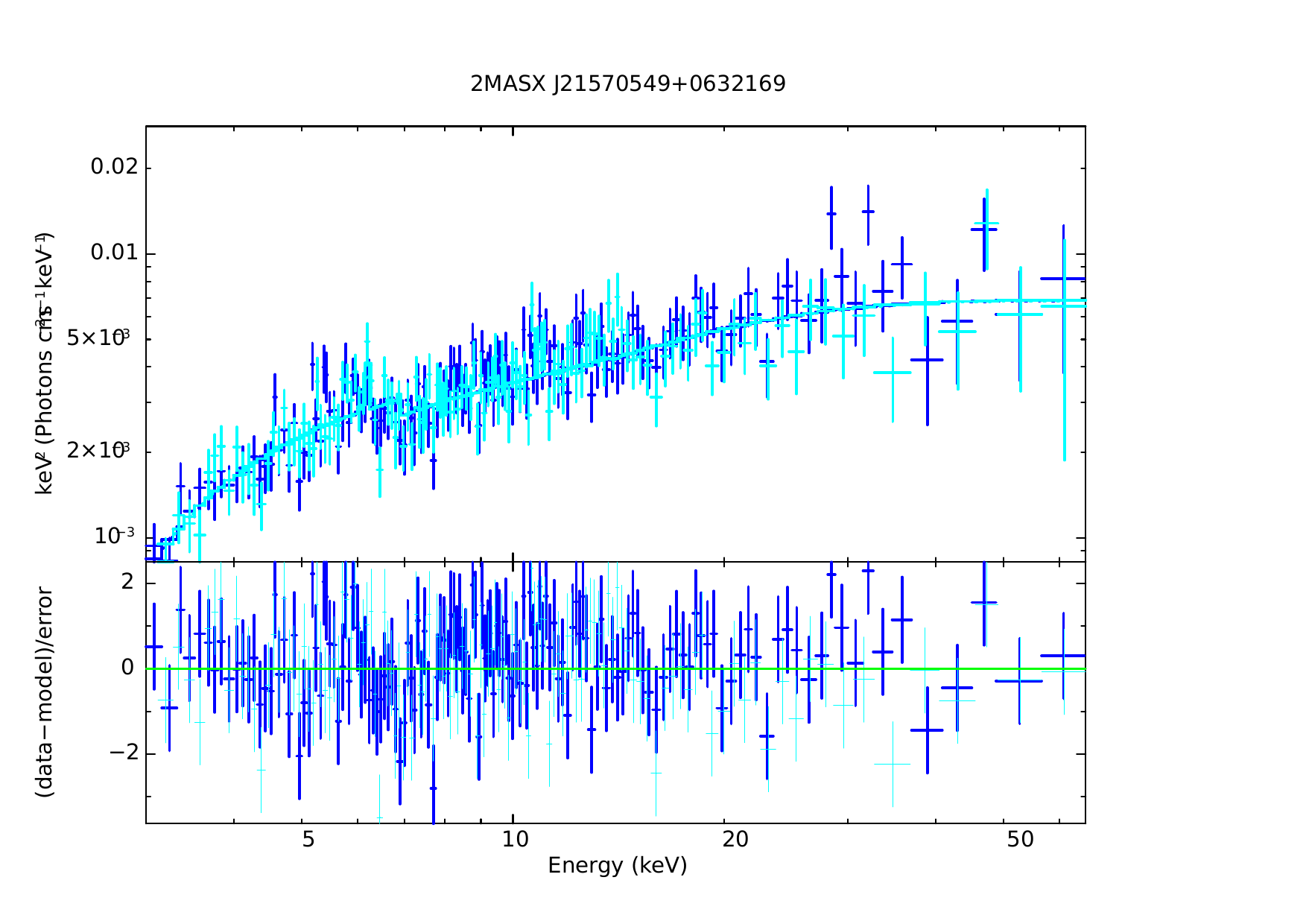}
    \caption{\uxc best fit of 2MASX J21570549$+$0632169.}
    \label{fig:2m21}
\end{figure*}

\begingroup
\renewcommand*{\arraystretch}{2}
\begin{table*}
    \centering
    \label{tab:sdss07}
    \begin{tabular}{lccc}
    \textbf{SDSS J073323.77$+$441448.8} \\
    \hline \hline
    \textbf{Model} & \textbf{UXClumpy} & \textbf{RXTorusD} \\
    \hline
     $\chi^2$/dof & 84/99 & 92/104 \\
    kT & ... & ... \\
    $\Gamma$ & 1.94$^{+0.14}_{-0.23}$ & 1.53$^{+0.12}_{-0.10}$ \\
    norm $10^{-2}$ & 0.17$^{+0.09}_{-0.07}$ & 0.05$^{+0.03}_{-0.02}$ \\
    CTKcover & 0.6$^{+u}_{-0.13}$ & ... \\
    Tor $\sigma$ & 45* & ... \\
    $\theta_{obs}$ & 45* & 90* \\
    r / R & ... & 0.5* \\
    $N_{H, l.o.s.}$ & 1.25$^{+0.40}_{-0.32}$ & 1.45$^{+0.35}_{-0.31}$ \\
    $N_{H, eq.}$ & ... & Fixed to $N_{H, l.o.s.}$ \\
    $f_s$ 10$^{-2}$ & 1* & 1* \\
    c$_{nus}$ & ... & ... \\
    \hline
    F$\rm_{2-10\,keV}$ & 1.58$^{+0.07}_{-0.07}$ $\times$ 10$^{-12}$ & 1.62$^{+0.08}_{-0.07}$ $\times$ 10$^{-12}$\\
    F$\rm_{15-55\,keV}$ & 3.89$^{+0.09}_{-0.18}$ $\times$ 10$^{-12}$ & 3.87$^{+0.18}_{-0.09}$ $\times$ 10$^{-12}$ \\
    L$\rm_{2-10\,keV}$ & 2.59$^{+1.42}_{-1.05}$ $\times$ 10$^{43}$ & 1.13$^{+0.27}_{-4.37}$ $\times$ 10$^{43}$ \\
    L$\rm_{15-55\,keV}$ & 2.95$^{+1.63}_{-1.19}$ $\times$ 10$^{43}$ & 2.85$^{+0.67}_{-1.08}$ $\times$ 10$^{43}$ \\
    \hline \hline
    \end{tabular}
\end{table*}
\endgroup

\iffalse
\begingroup
\renewcommand*{\arraystretch}{2}
\begin{table*}
    \centering
    \label{tab:sdss07}
    \begin{tabular}{lccc}
    \textbf{SDSS J073323.77$+$441448.8} \\
    \hline \hline
    \textbf{Model} & \textbf{UXClumpy} & \textbf{RXTorusD} \\
    \hline
     $\chi^2$/dof & 84/99 & 96/104 \\
    kT & ... & ... \\
    $\Gamma$ & 1.94$^{+0.14}_{-0.23}$ & 1.64$^{+0.12}_{-0.12}$ \\
    norm $10^{-2}$ & 0.17$^{+0.09}_{-0.07}$ & 0.06$^{+0.03}_{-0.02}$ \\
    CTKcover & 0.6$^{+u}_{-0.13}$ & ... \\
    Tor $\sigma$ & 45* & ... \\
    $\theta_{obs}$ & 45* & 45* \\
    r / R & ... & 0.5* \\
    $N_{H, l.o.s.}$ & 1.25$^{+0.40}_{-0.32}$ & 1.16$^{+0.04}_{-0.03}$ \\
    $N_{H, eq.}$ & ... & Fixed to $N_{H, l.o.s.}$ \\
    $f_s$ 10$^{-2}$ & 1* & 1* \\
    c$_{nus}$ & ... & ... \\
    \hline
    F$\rm_{2-10\,keV}$ & 1.58$^{+0.07}_{-0.07}$ $\times$ 10$^{-12}$ & 1.62$^{+0.08}_{-0.07}$ $\times$ 10$^{-12}$\\
    F$\rm_{15-55\,keV}$ & 3.89$^{+0.09}_{-0.18}$ $\times$ 10$^{-12}$ & 3.87$^{+0.18}_{-0.09}$ $\times$ 10$^{-12}$ \\
    L$\rm_{2-10\,keV}$ & 2.59$^{+1.42}_{-1.05}$ $\times$ 10$^{43}$ & 1.13$^{+0.27}_{-4.37}$ $\times$ 10$^{43}$ \\
    L$\rm_{15-55\,keV}$ & 2.95$^{+1.63}_{-1.19}$ $\times$ 10$^{43}$ & 2.85$^{+0.67}_{-1.08}$ $\times$ 10$^{43}$ \\
    \hline \hline
    \end{tabular}
\end{table*}
\endgroup
\fi

\begin{figure*}
    \centering
    \includegraphics[angle=270, scale=0.6]{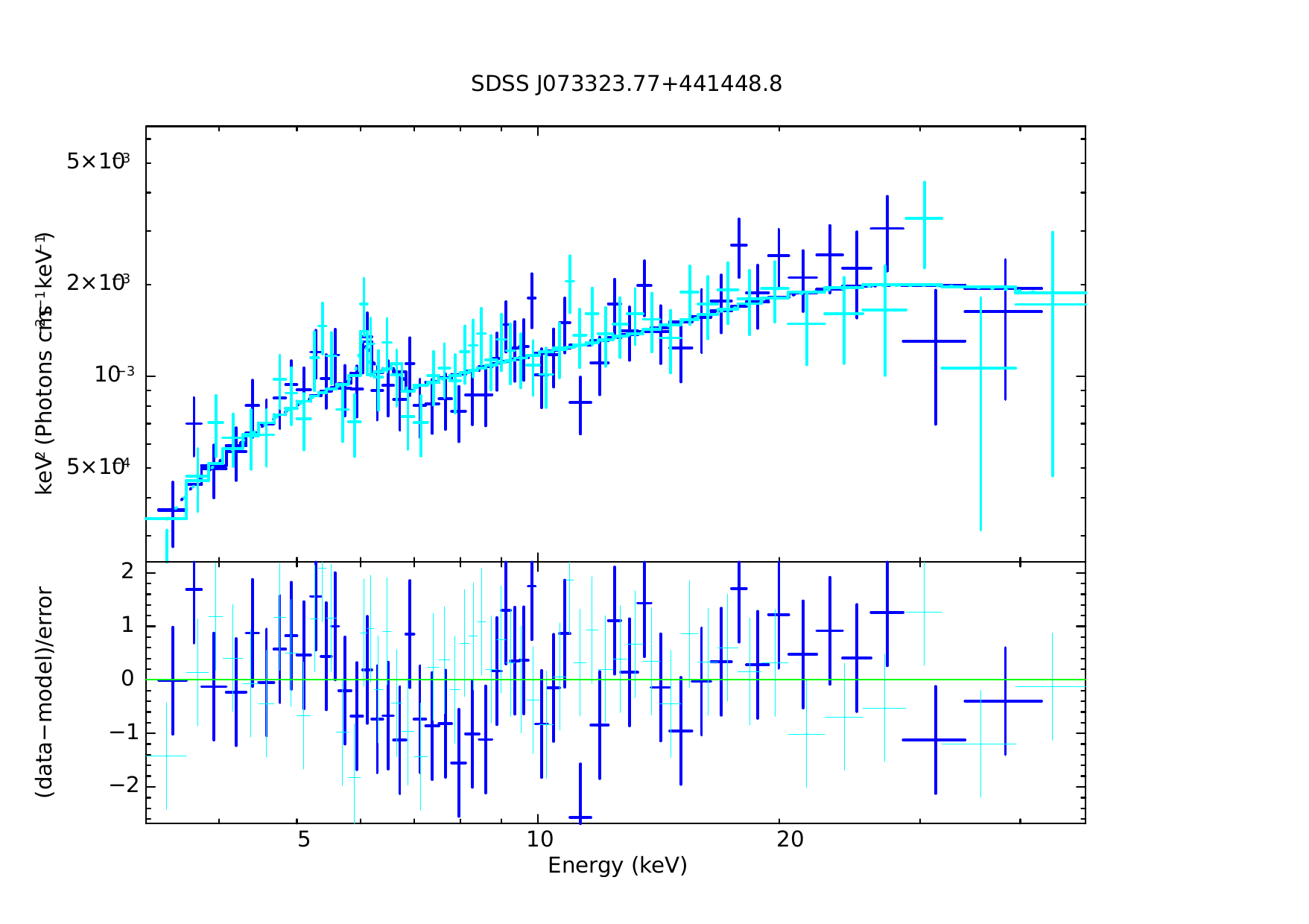}
    \caption{\uxc best fit of SDSS J073323.77$+$441448.8.}
    \label{fig:sdss}
\end{figure*}

\begingroup
\renewcommand*{\arraystretch}{2}
\begin{table*}
    \centering
    \label{tab:2m13}
    \begin{tabular}{lccc}
    \textbf{2MASX J13100789$-$1711398} \\
    \hline \hline
    \textbf{Model}  & \textbf{UXClumpy} & \textbf{RXTorusD} \\
    \hline
    $\chi^2$/dof & 178 / 160 & 147 / 159 \\
    kT &  0.25$^{+0.22}_{-0.07}$ & 0.29$^{+0.41}_{-0.10}$ \\
    $\Gamma$ & 1.89$^{+0.10}_{-0.11}$ & 1.95$^{+0.19}_{-0.21}$  \\
    norm $10^{-2}$ & 0.18$^{+0.06}_{-0.05}$ & 0.20$^{+0.14}_{-0.09}$ \\
    CTKcover & 0$^{+0.14}_{-u}$ & ... \\
    Tor $\sigma$ & 84$^{+u}_{-6}$ & ... \\
    $\theta_{obs}$ & 45* & 15.4$^{+4.8}_{-3.3}$ \\
    r / R & ... & 0.97$^{+0.02}_{-0.02}$ \\
    $N_{H, l.o.s.}$ & 3.58$^{+0.35}_{-0.39}$ & 8.67$^{+2.65}_{-2.14}$ \\
    $N_{H, eq.}$ & ... & 6.31$^{+2.55}_{-2.44}$ \\
    $f_s$ 10$^{-2}$ & 9.9$^{+4.1}_{-2.6}$ & 0.67$^{+0.60}_{-0.40}$ \\
    c$_{nus}$ & 1.19$^{+0.09}_{-0.09}$ & 1.21$^{+0.09}_{-0.09}$  \\
    \hline
    F$\rm_{2-10\,keV}$ & 1.14$^{+0.03}_{-0.05}$ $\times$ 10$^{-12}$ & 1.15$^{+0.03}_{-0.03}$ $\times$ 10$^{-12}$ \\
    F$\rm_{15-55\,keV}$ & 1.97$^{+0.09}_{-0.09}$ $\times$ 10$^{-13}$ & 2.01$^{+0.09}_{-0.13}$ $\times$ 10$^{-13}$ \\
    L$\rm_{2-10\,keV}$ & 7.19$^{+2.36}_{-2.02}$ $\times$ 10$^{42}$ & 9.57$^{+1.13}_{-0.70}$ $\times$ 10$^{42}$\\
    L$\rm_{15-55\,keV}$ & 8.02$^{+2.05}_{-2.24}$ $\times$ 10$^{42}$ & 1.22$^{+0.57}_{-0.35}$ $\times$ 10$^{43}$ \\
    \hline \hline
    \end{tabular}
\end{table*}
\endgroup

\begin{figure*}
    \centering
    \includegraphics[angle=270, scale=0.6]{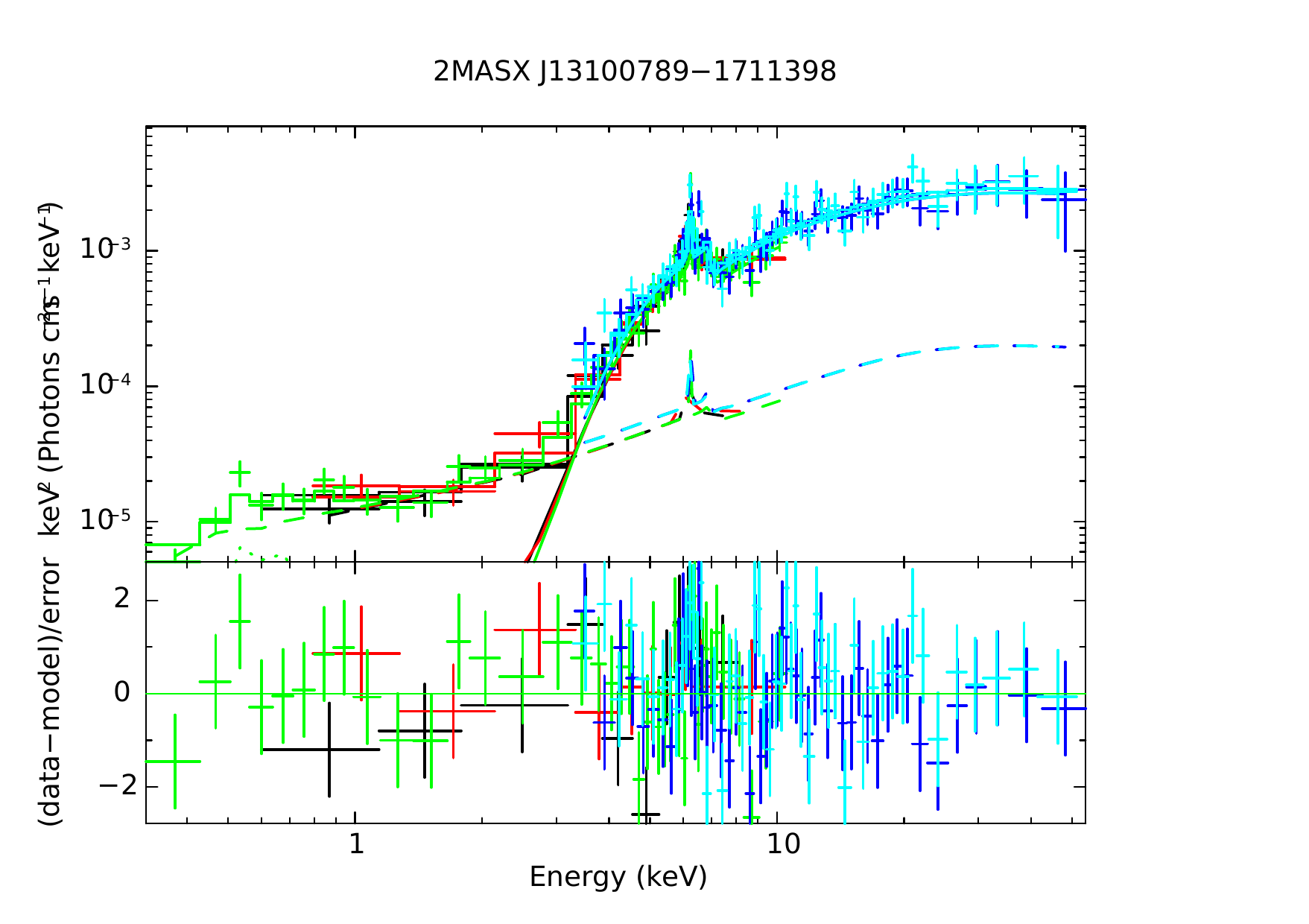}
    \caption{UXClumpy best fit of 2MASX J13100789$-$1711398.}
    \label{fig:2m1310}
\end{figure*}

\begingroup
\renewcommand*{\arraystretch}{2}
\begin{table*}
    \centering
    \label{tab:leda20}
    \begin{tabular}{lccc}
    \textbf{LEDA 2019751} \\
    \hline \hline
    \textbf{Model}  & \textbf{UXClumpy} & \textbf{RXTorusD} \\
    \hline
    $\chi^2$/dof & 413 / 374 & 419 / 375 \\
    kT &  0.6* &  0.6* \\
    $\Gamma$ & 1.79$^{+0.07}_{-0.06}$ & 1.70$^{+0.05}_{-0.06}$ \\
    norm $10^{-2}$ & 0.13$^{+0.02}_{-0.02}$ & 0.07$^{+0.01}_{-0.01}$ \\
    CTKcover & 0$^{+0.15}_{-u}$ & ... \\
    Tor $\sigma$ & 11$^{+41}_{-7}$ & ... \\
    $\theta_{obs}$ & 90$^{+u}_{-44}$ & 3.0$^{+13.2}_{-u}$ \\
    r / R & ... & 1.0$^{+u}_{-0.88}$ \\
    $N_{H, l.o.s.}$ & 0.52$^{+0.04}_{-0.04}$ & 0.65$^{+0.05}_{-0.03}$ \\
    $N_{H, eq.}$ & ... & 1.33$^{+0.87}_{-0.73}$ \\
    $f_s$ 10$^{-2}$ & 1.0* & 1.0* \\
    c$_{nus}$ & 1.17$^{+0.05}_{-0.05}$ & 1.15$^{+0.05}_{-0.04}$ \\
    \hline
    F$\rm_{2-10\,keV}$ & 2.34$^{+0.05}_{-0.05}$ $\times$ 10$^{-12}$ & 2.33$^{+0.02}_{-0.07}$ $\times$ 10$^{-12}$\\
    F$\rm_{15-55\,keV}$ & 1.78$^{+0.04}_{-0.08}$ $\times$ 10$^{-13}$ & 1.80$^{+0.09}_{-0.04}$ $\times$ 10$^{-13}$ \\
    L$\rm_{2-10\,keV}$ & 5.33$^{+0.82}_{-0.82}$ $\times$ 10$^{43}$ & 5.16$^{+0.27}_{-0.66}$ $\times$ 10$^{43}$ \\
    L$\rm_{15-55\,keV}$ & 6.94$^{+1.06}_{-1.08}$ $\times$ 10$^{43}$ & 8.03$^{+0.59}_{-1.47}$ $\times$ 10$^{43}$\\
    \hline \hline
    \end{tabular}
\end{table*}
\endgroup

\begin{figure*}
    \centering
    \includegraphics[angle=270, scale=0.6]{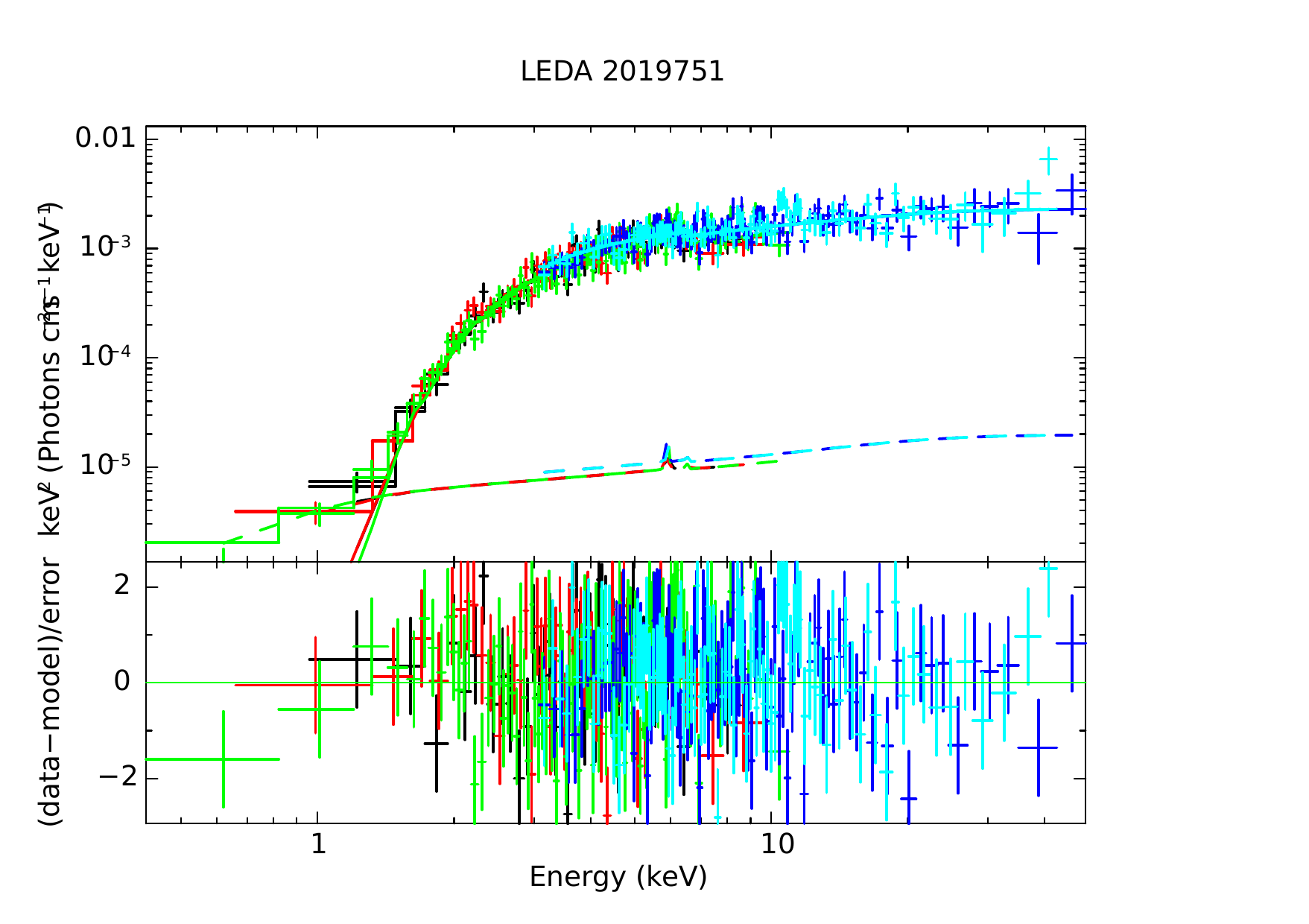}
    \caption{UXClumpy best fit of LEDA 2019751.}
    \label{fig:leda2019}
\end{figure*}

\begingroup
\renewcommand*{\arraystretch}{2}
\begin{table*}
    \centering
    \label{tab:ngc52}
    \begin{tabular}{lccc}
    \textbf{NGC 52} \\
    \hline \hline
    \textbf{Model}  & \textbf{UXClumpy} & \textbf{RXTorusD} \\
    \hline
    $\chi^2$/dof & 370 / 351 & 367 / 352 \\
    kT &  0.6* & 0.6* \\
    $\Gamma$ & 1.85$^{+0.07}_{-0.07}$ & 1.51$^{+0.04}_{-0.05}$ \\
    norm $10^{-2}$ & 0.17$^{+0.04}_{-0.02}$ & 0.05$^{+0.01}_{-0.01}$ \\
    CTKcover & 0$^{+0.28}_{-u}$ & ... \\
    Tor $\sigma$ & 42$^{+31}_{-16}$ & ... \\
    $\theta_{obs}$ & 0$^{+68}_{-u}$ & 25.8$^{+22.9}_{-5.8}$ \\
    r / R & ... & 1.0$^{+u}_{-0.10}$ \\
    $N_{H, l.o.s.}$ & 1.17$^{+0.08}_{-0.09}$ & 1.21$^{+0.21}_{-0.11}$ \\
    $N_{H, eq.}$ & ... & 2.52$^{+0.64}_{-1.00}$ \\
    $f_s$ 10$^{-2}$ & 1.0* & 1.0* \\
    c$_{nus}$ & 1.21$^{+0.05}_{-0.05}$ & 1.21$^{+0.05}_{-0.05}$ \\
    \hline
    F$\rm_{2-10\,keV}$ & 2.31$^{+0.05}_{-0.05}$ $\times$ 10$^{-12}$ & 2.30$^{+0.06}_{-0.06}$ $\times$ 10$^{-12}$\\
    F$\rm_{15-55\,keV}$ & 2.37$^{+0.11}_{-0.05}$ $\times$ 10$^{-13}$ & 2.32$^{+0.10}_{-0.061}$ $\times$ 10$^{-13}$ \\
    L$\rm_{2-10\,keV}$ & 3.15$^{+0.76}_{-0.36}$ $\times$ 10$^{42}$ & 2.44$^{+0.28}_{-0.31}$ $\times$ 10$^{42}$\\
    L$\rm_{15-55\,keV}$ & 3.84$^{+0.93}_{-0.43}$ $\times$ 10$^{42}$ & 5.45$^{+0.94}_{-1.06}$ $\times$ 10$^{42}$ \\
    \hline \hline
    \end{tabular}
\end{table*}
\endgroup

\begin{figure*}
    \centering
    \includegraphics[angle=270, scale=0.6]{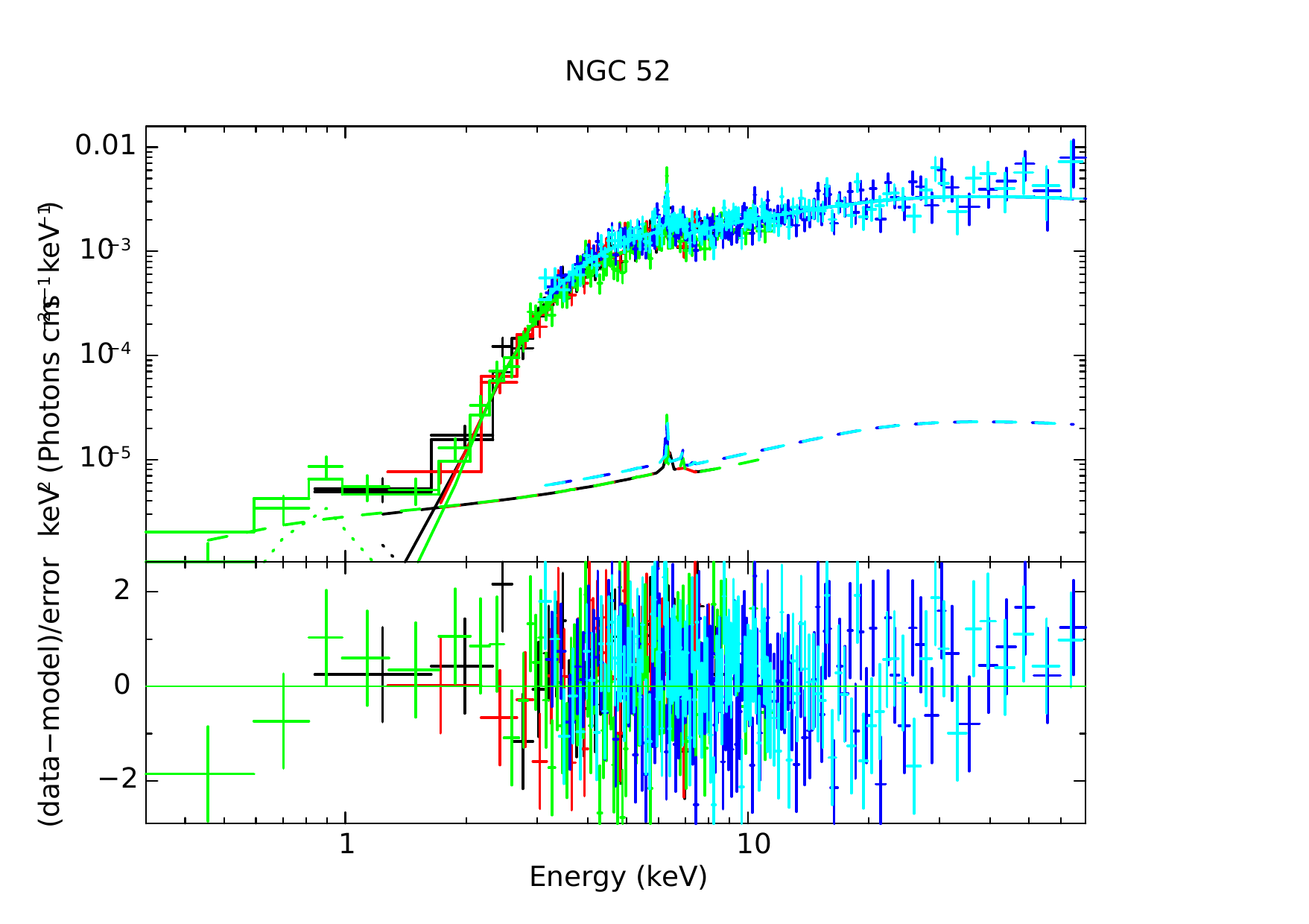}
    \caption{UXClumpy best fit of NGC 52.}
    \label{fig:ngc52}
\end{figure*}

\begingroup
\renewcommand*{\arraystretch}{2}
\begin{table*}
    \centering
    \caption{\textbf{Notes:} The 15$-$55\,keV fluxes listed are upper limits.}
    \label{tab:ugc12}
    \begin{tabular}{lccc}
    \textbf{UGC 12566} \\
    \hline \hline
    \textbf{Model}  & \textbf{UXClumpy} & \textbf{RXTorusD} \\
    \hline
    $\chi^2$/dof & 42 / 55 & 43 / 55\\
    kT &  0.32$^{+0.51}_{-0.09}$ &  0.29$^{+0.43}_{-0.09}$ \\
    $\Gamma$ & 2.09$^{+0.32}_{-0.34}$ & 1.77$^{+0.34}_{-0.34}$ \\
    norm $10^{-2}$ & 0.03$^{+0.03}_{-0.02}$ & 0.01$^{+0.04}_{-0.01}$ \\
    CTKcover & 0.4* & ... \\
    Tor $\sigma$ & 45* & ... \\
    $\theta_{obs}$ & 45* & 90* \\
    r / R & ... & 0.5* \\
    $N_{H, l.o.s.}$ & 1.59$^{+0.40}_{-0.37}$ & 2.13$^{+0.46}_{-0.49}$ \\
    $N_{H, eq.}$ & ... & Fixed to $N_{H, l.o.s.}$  \\
    $f_s$ 10$^{-2}$ & 5.48$^{+5.10}_{-2.80}$ & 1.0* \\
    c$_{nus}$ & 1.25$^{+0.27}_{-0.24}$ & 1.26$^{+0.28}_{-0.24}$ \\
    \hline
    F$\rm_{2-10\,keV}$ & 1.95$^{+0.19}_{-0.13}$ $\times$ 10$^{-13}$ & 1.86$^{+0.14}_{-0.13}$ $\times$ 10$^{-13}$\\
    F$\rm_{15-55\,keV}$ & 1.62$^{+0.16}_{-0.19}$ $\times$ 10$^{-14}$ & 1.55$^{+0.14}_{-0.17}$ $\times$ 10$^{-14}$ \\
    L$\rm_{2-10\,keV}$ & 4.24$^{+4.93}_{-2.33}$ $\times$ 10$^{41}$ & 1.46$^{+3.05}_{-0.71}$ $\times$ 10$^{41}$ \\
    L$\rm_{15-55\,keV}$ & 4.87$^{+4.91}_{-2.69}$ $\times$ 10$^{41}$ & 3.20$^{+6.70}_{-1.55}$ $\times$ 10$^{41}$ \\
    \hline \hline
    \end{tabular}
\end{table*}
\endgroup

\iffalse
\begingroup
\renewcommand*{\arraystretch}{2}
\begin{table*}
    \centering
    \caption{\textbf{Notes:} The 15$-$55\,keV fluxes listed are upper limits.}
    \label{tab:ugc12}
    \begin{tabular}{lccc}
    \textbf{UGC 12566} \\
    \hline \hline
    \textbf{Model}  & \textbf{UXClumpy} & \textbf{RXTorusD} \\
    \hline
    $\chi^2$/dof & 42 / 55 & 42 / 55\\
    kT &  0.32$^{+0.51}_{-0.09}$ &  0.30$^{+0.43}_{-0.09}$ \\
    $\Gamma$ & 2.09$^{+0.32}_{-0.34}$ & 2.03$^{+0.39}_{-0.38}$ \\
    norm $10^{-2}$ & 0.03$^{+0.03}_{-0.02}$ & 0.02$^{+0.04}_{-0.01}$ \\
    CTKcover & 0.4* & ... \\
    Tor $\sigma$ & 45* & ... \\
    $\theta_{obs}$ & 45* & 45* \\
    r / R & ... & 0.5* \\
    $N_{H, l.o.s.}$ & 1.59$^{+0.40}_{-0.37}$ & 2.02$^{+0.59}_{-0.49}$ \\
    $N_{H, eq.}$ & ... & Fixed to $N_{H, l.o.s.}$  \\
    $f_s$ 10$^{-2}$ & 5.48$^{+5.10}_{-2.80}$ & 1.0* \\
    c$_{nus}$ & 1.25$^{+0.27}_{-0.24}$ & 1.26$^{+0.28}_{-0.24}$ \\
    \hline
    F$\rm_{2-10\,keV}$ & 1.95$^{+0.19}_{-0.13}$ $\times$ 10$^{-13}$ & 1.86$^{+0.14}_{-0.13}$ $\times$ 10$^{-13}$\\
    F$\rm_{15-55\,keV}$ & 1.62$^{+0.16}_{-0.19}$ $\times$ 10$^{-14}$ & 1.55$^{+0.14}_{-0.17}$ $\times$ 10$^{-14}$ \\
    L$\rm_{2-10\,keV}$ & 4.24$^{+4.93}_{-2.33}$ $\times$ 10$^{41}$ & 1.46$^{+3.05}_{-0.71}$ $\times$ 10$^{41}$ \\
    L$\rm_{15-55\,keV}$ & 4.87$^{+4.91}_{-2.69}$ $\times$ 10$^{41}$ & 3.20$^{+6.70}_{-1.55}$ $\times$ 10$^{41}$ \\
    \hline \hline
    \end{tabular}
\end{table*}
\endgroup
\fi

\begin{figure*}
    \centering
    \includegraphics[angle=270, scale=0.6]{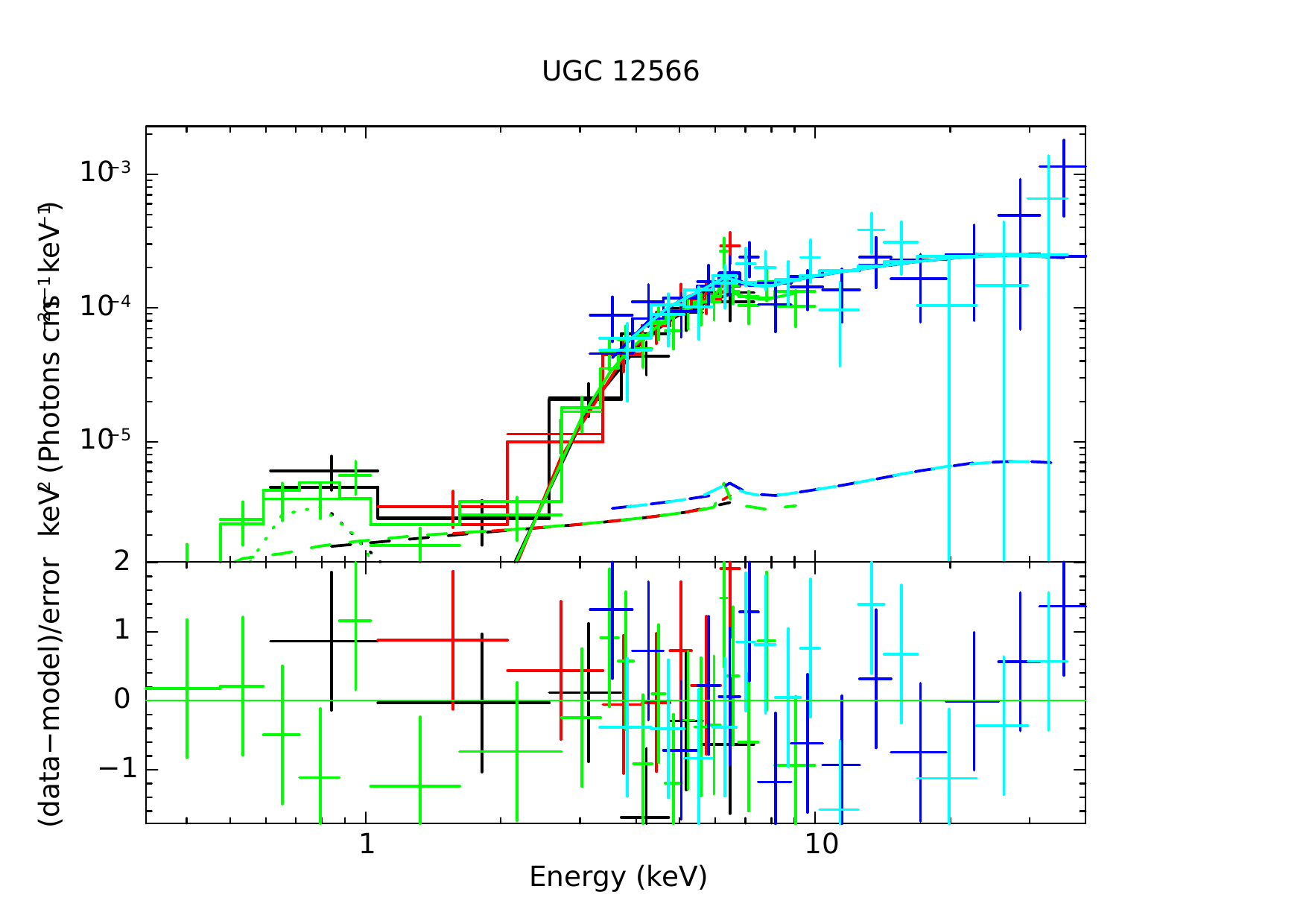}
    \caption{UXClumpy best fit of UGC 12566.}
    \label{fig:ugc12}
\end{figure*}

\begingroup
\renewcommand*{\arraystretch}{2}
\begin{table*}
    \centering
    \label{tab:ngc6657}
    \begin{tabular}{lccc}
    \textbf{NGC 6657} \\
    \hline \hline
    \textbf{Model}  & \textbf{UXClumpy} & \textbf{RXTorusD} \\
    \hline
    $\chi^2$/dof & 319 / 299 & 318 / 295 \\
    kT &  0.0* &  0.0* \\
    $\Gamma$ & 1.88$^{+0.07}_{-0.10}$ & 1.61$^{+0.12}_{-0.10}$ \\
    norm $10^{-2}$ & 0.18$^{+0.02}_{-0.03}$ & 0.07$^{+0.03}_{-0.03}$ \\
    CTKcover & 0.20$^{+0.24}_{-0.03}$ & ... \\
    Tor $\sigma$ & 21.2$^{+u}_{-3.1}$ & ... \\
    $\theta_{obs}$ & 45* & 50.0$^{+u}_{-19.0}$ \\
    r / R & ... & 0.95$^{+u}_{-0.34}$ \\
    $N_{H, l.o.s.}$ & 1.26$^{+0.06}_{-0.18}$ & 2.25$^{+2.15}_{-0.72}$ \\
    $N_{H, eq.}$ & ... & 1.00$^{+u}_{-0.29}$ \\
    $f_s$ 10$^{-2}$ & 1* & 0.75$^{+0.52}_{-0.31}$ \\
    c$_{nus}$ & 1.20$^{+0.06}_{-0.06}$ & 1.19$^{+0.06}_{-0.05}$ \\
    \hline
    F$\rm_{2-10\,keV}$ & 1.98$^{+0.05}_{-0.05}$ $\times$ 10$^{-12}$ & 2.00$^{+0.03}_{-0.06}$ $\times$ 10$^{-12}$\\
    F$\rm_{15-55\,keV}$ & 1.97$^{+0.09}_{-0.05}$ $\times$ 10$^{-13}$ & 1.95$^{+0.04}_{-0.09}$ $\times$ 10$^{-13}$ \\
    L$\rm_{2-10\,keV}$ & 5.34$^{+0.64}_{-0.85}$ $\times$ 10$^{42}$ & 4.91$^{+0.82}_{-1.02}$ $\times$ 10$^{42}$ \\
    L$\rm_{15-55\,keV}$ & 6.97$^{+0.83}_{-1.12}$ $\times$ 10$^{42}$ & 8.30$^{+2.57}_{-3.22}$ $\times$ 10$^{42}$\\
    \hline \hline
    \end{tabular}
\end{table*}
\endgroup

\begin{figure*}
    \centering
    \includegraphics[angle=270, scale=0.6]{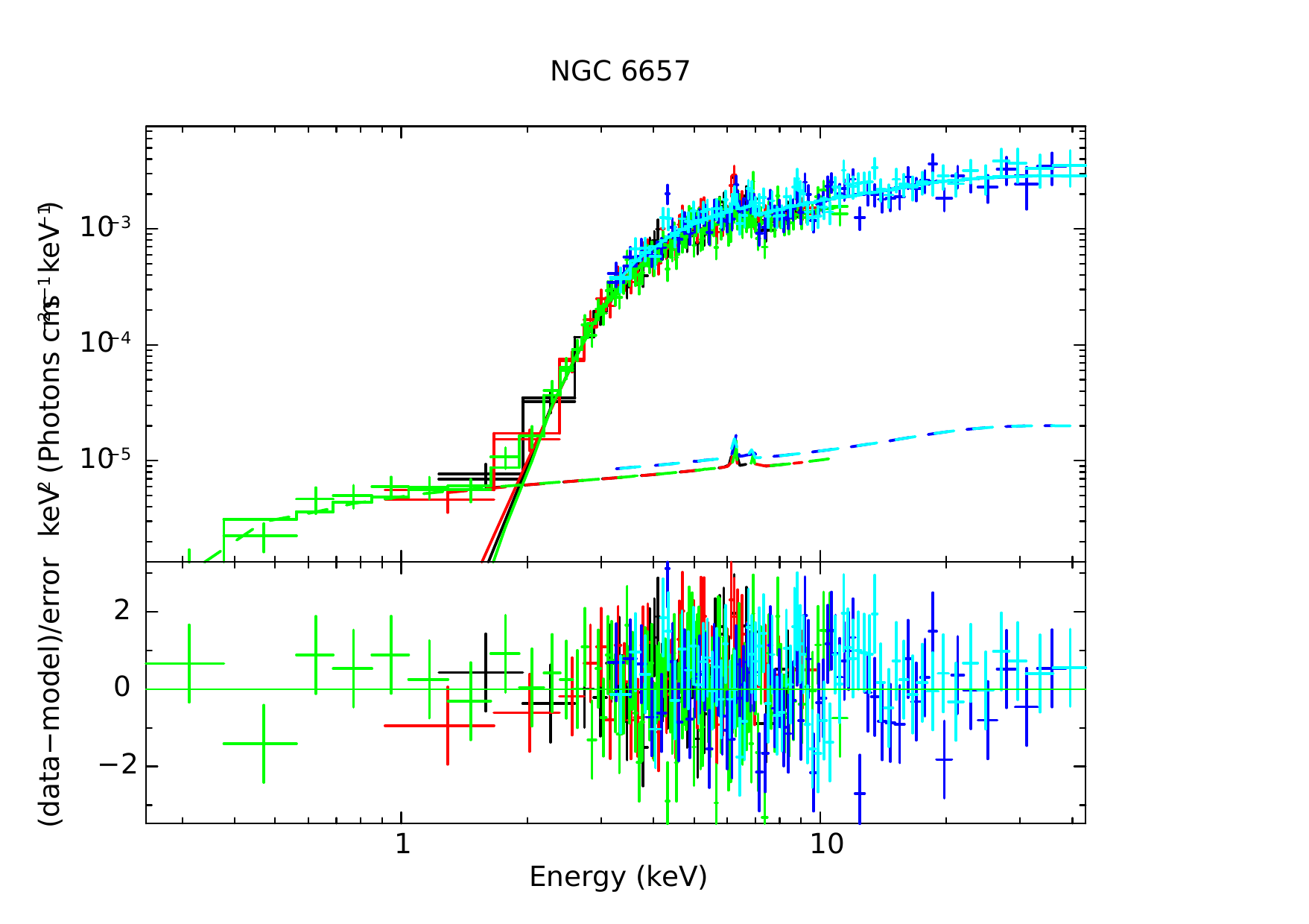}
    \caption{UXClumpy best fit of NGC 6657.}
    \label{fig:ngc6657}
\end{figure*}

%\section{Extra Plots}

\iffalse
\begin{figure*}
    \centering
    \includegraphics[scale=0.8]{images/nh_hist.png}
    \caption{Histogram of line-of-sight column density values found with \texttt{UXClumpy} (red) and \texttt{RXTorusD} (blue). For source 2MASX J13100789$-$1711398, I used the MYTorus NH instead of the RXTorusD NH.}
    \label{fig:nh_hist}
\end{figure*}
\fi

\bibliographystyle{aasjournal}
\bibliography{bibliography}

\end{document}